\documentclass[aps,a4paper,superscriptaddress,nofootinbib,12pt]{revtex4}
\usepackage[T1]{fontenc}
\usepackage{hyperref}
\usepackage{amsmath}
\usepackage{amsthm}
\usepackage{amsfonts}
\usepackage{xcolor}
\usepackage{amssymb}
\usepackage{stmaryrd}
\usepackage[toc,page]{appendix}
\newtheorem{theorem}{Theorem}[section]
\newtheorem{definition}{Definition}[section]

\def\ca{{\cal A}}
\def\cb{{\cal B}}
\def\cc{{\cal C}}

\def\cv{{\cal V}}

\begin{document}
\title{Palatini frames in scalar-tensor theories of gravity
}
\author{Aleksander Kozak}
\email{aleksander.kozak@ift.uni.wroc.pl}
\affiliation{Institute of Theoretical Physics, University of Wroclaw, pl. M. Borna 9,
50-204 Wroclaw, Poland.}
 
\author{Andrzej Borowiec}
\email{andrzej.borowiec@ift.uni.wroc.pl}
\affiliation{Institute of Theoretical Physics, University of Wroclaw, pl. M. Borna 9,
50-204 Wroclaw, Poland.}

\begin{abstract}
A new systematic approach extending the notion of frames to the  Palatini scalar-tensor theories of gravity in various dimensions $n>2$ is proposed. We impose frame transformation induced by the group action which includes almost-geodesic and conformal transformations. We characterize theories invariant with respect to these transformations dividing them up into solution-equivalent subclasses (group orbits). To this end, invariant characteristics have been introduced. The formalism provides new frames incorporating non-metricity that lead to re-definition of Jordan frames.
The case of Palatini $F(R)$-gravity is considered in more detail.
\end{abstract}

\maketitle
\tableofcontents

\section{Introduction}
	Despite many theoretical and experimental triumphs \cite{test}, including recent detection of gravitational waves \cite{Abbott}, general relativity is not considered a fundamental theory describing gravitational interactions; see e.g. \cite{NO06}--\cite{nodi}. Based on our current understanding of the workings of Nature, a few arguments for modifying it can be given. First of all, GR cannot be satisfactorily quantized, as attempts to renormalize it have been futile. Secondly, it is not a low-energy limit of theories regarded as fundamental, such as bosonic string theories \cite{string}, where dilaton fields couple non-minimally to the spacetime curvature.
	Another problem concerns the $\Lambda$CDM model: it is customary to consider that the value of $\Lambda$ being responsible for the current acceleration of the expansion of the Universe is usually incomprehensibly small (120 order of magnitude smaller) when compared to the value predicted by quantum field theory. In fact, more realistic estimations taking into account   Pauli-Zeldovich cancellation effect, quantum field theory in curved background or supersymmetry, make this discrepancy not so drastic  (for more discussion see \cite{staro}- \cite{Kamenshchik}).

	As far as the mathematical reasons for modifying the Einstein's gravity are concerned, we can take the so-called Palatini formalism into consideration. In the standard gravity, the underlying assumption of geometric structures defined on spacetime is that the affine connection is the Levi-Civita connection of the metric. In the Palatini approach, however, we consider these two objects as unrelated, since there is no reason whatsoever we should impose a relation between them a priori. In case of Einstein gravity, introducing Palatini formalism does not affect the resulting field equations in any way; however, in case of more complicated theories, such as scalar-tensor or $F(R)$ theories of gravity, both approaches usually give different results, describing different physics. Palatini formalism has been investigated especially in the context of cosmological applications \cite{fund}-\cite{ssb}.
	
	Scalar-tensor (S-T) theories of gravity are a very promising modification of the Einstein gravity. In these theories, a scalar field is non-minimally coupled to the curvature scalar \cite{fuj}. Historically, the prototype of all contemporary scalar-tensor theories was the Brans-Dicke theory \cite{dicke}.  An interesting feature of the scalar-tensor theories of gravity is their equivalence to the $F(R)$ theories, which basically means that the latter can be analyzed using the "mathematical machinery" developed for the former \cite{Sotiriou:2006hs}. The reason why the scalar-tensor theories deserve some attention is that they can be successfully used to build credible models for cosmic inflation \cite{starobinsky2}  (utilizing the equivalence between the scalar-tensor and $F(R)$ theories of gravity) and dark energy \cite{jarv}.
	
	Hitherto, the scalar-tensor theories of gravity have been considered mostly in a purely metric approach \cite{fund}, \cite{fuj}, \cite{jarv}-\cite{Karam2} and the possible effects of adopting the Palatini approach have been analyzed somewhat less commonly \cite{Ulf}-\cite{vecthorn}. So far, general conditions for a correct formulation of the scalar-tensor theories have been analyzed \cite{hownot}.  Change of formalism from metric to Palatini applied to S-T theories has been investigated in the context of cosmology, to analyze the problem of cosmological constant \cite{bauer2}, quintessence - to show that equation of state in the Palatini formalism can cross the phantom divide line \cite{wang}, and inflation, where it was discovered that in the Palatini approach \cite{racc}-\cite{tommi}, inflationary epoch is naturally provied \cite{racc}-\cite{bauer}, and almost scale-invariant curvature perturbations are generated with no tensor modes \cite{tamanini}. Some authors generalized scalar-tensor theories and allowed non-minimal derivative coupling as well \cite{dmitri}-\cite{luo}. In such theories, one makes extensive use of 
so-called "disformal transformations". It was shown that for a special choice of parameters characterizing the theory, adopting Palatini approach allows one to avoid Ostrogradski ghosts \cite{dmitri}  \footnote{ It should be noted that the disformal transformations can be combined together with the conformal transformations considered in the present  paper, see e.g. \cite{ ST14}.}.  Also, vector-Horndeski theories were analyzed with the metric structure decoupled from the affine structure. It was proven that in the Palatini formalism, there exist cosmological solutions which can pass through singularities  \cite{vecthorn}. 
	
	The main goal of this paper is to introduce the general theory of scalar-tensor gravity analyzed in the Palatini approach and to develop mathematical formalism enabling us to analyze any S-T theory in a (conformally) frame-independent manner. The outline of this paper 
	\footnote{This is an extension of the results obtained initially in \cite{Kozak}.} goes as follows: in the first part, postulated action functional will be presented, and equations of motion derived. Next, modified conformal transformations in the Palatini approach will be introduced in order to allow the connection to transform independently of the metric tensor. A solution of the equation resulting from varying with respect to the independent connection will be inspected. Then, following the procedure carried out in \cite{jarv} (see also \cite{flan}, \cite{Karam}), invariant quantities defined for the Palatini S-T theory will be obtained. The results will be applied to an analysis of $F(R)$ Palatini gravity. In the last part, general conditions on the possible equivalence between a given S-T theory and some $F(R)$ gravity will be discussed. For reader's convenience, some supplementary material is collected in four Appendices.
	
	 	\section{Action functional and equations of motion} 
	The main idea behind the Palatini approach is the following: we no longer consider metric tensor and linear connection to be dependent on each other. This approach was originally analyzed by Einstein \cite{ein}, but then was attributed to an Italian mathematician Attilio Palatini \cite{Palatini,pal}. In this approach, one decouples causal structure of spacetime from its affine structure (which determines geodesics followed by particles).
 In practical terms, Palatini formalism amounts to varying the action functional with respect to both the metric tensor and the torsionless (i.e. symmetric) affine connection, resulting in two sets of field equations. One of these sets establishes a relation between the metric and the connection. There is no particular reason to apply the Palatini variation to the standard Einstein-Hilbert action, as in that case the independent connection turns out to be Levi-Civita with respect to the metric tensor, i.e. related to the metric by the standard formula: $\Gamma^\alpha_{\mu\nu}=\frac{1}{2}g^{\alpha\beta}(\partial_\mu g_{\beta\nu}+\partial_\nu g_{\beta\mu}-\partial_\beta g_{\mu\nu})$. 
However, in case of more complicated theories, such as $F(R)$ theories of gravity, where the curvature scalar in the Einstein-Hilbert action is replaced by a function of it, both approaches give physically incompatible results, leading to different field equations describing different physics in the presence of matter sources. Instead, in  the vacuum case, the Einstein equations enriched by adding cosmological constant are still valid \cite{uni1}, \cite{uni2}.
	
Consider a triple $(M, \Gamma, g)$, where $M$ is $n$-dimensional $n>2$ manifold 
\footnote{For two-dimensional case see e.g. \cite{Ulf3}, \cite{uni1}.} equipped with a torsion-free ($\equiv$ symmetric) connection $\Gamma=\Gamma_{\mu\nu}^\alpha=\Gamma_{\nu\mu}^\alpha$ and a metric tensor $g=g_{\mu\nu}$,	possibly of the Lorentzian signature.  The affine  connection is used to build the Riemann curvature tensor:
	\begin{equation}
	R^\alpha_{\:\mu\beta\nu}(\Gamma)=\partial_\beta \Gamma^\alpha_{\mu\nu}-\partial_\nu\Gamma^\alpha_{\mu\beta}+\Gamma^\alpha_{\beta\sigma}\Gamma^\sigma_{\nu\mu}-\Gamma^\alpha_{\nu\sigma}\Gamma^\sigma_{\beta\mu}.
	\end{equation}
	
	The curvature scalar is a function of both the connection and the metric tensor:
	\begin{equation}
	R(g,\Gamma)=g^{\mu\nu}R_{\mu\nu}(\Gamma),
	\end{equation}
	where $R_{\mu\nu}(\Gamma)=R^\alpha_{\:\mu\alpha\nu}(\Gamma)$. 
	
Utilizing the Palatini approach, we want now to write down the most general action functional for scalar-tensor theories, which is consistent with some class of transformations (see explanations below and Appendix B). The action should contain a scalar field $\Phi$ - or a function thereof - non-minimally coupled to the curvature defined above and possibly to the matter fields. Furthermore, one must include also a kinetic term rendering the scalar field dynamic, and a self-interaction potential of the field. Presence of additional terms resulting from the approach we adopt, absent in the metric version of the theory, cannot be excluded.
	
	Therefore, we postulate the following action functional:
	\begin{equation}
	\begin{split}
	 S[g_{\mu\nu},\Gamma^\alpha_{\mu\nu},\Phi]=\frac{1}{2\kappa^2}&\int_{\Omega}d^nx\sqrt{-g}\Big[\mathcal{A}(\Phi)R(g,\Gamma)-\mathcal{B}(\Phi)g^{\mu\nu}\nabla_\mu\Phi\nabla_\nu\Phi-A_1^\mu(g,\Gamma)\mathcal{C}_1(\Phi)\nabla_\mu\Phi\\
	& -A_2^\mu(g,\Gamma)\mathcal{C}_2(\Phi)\nabla_\mu\Phi-\mathcal{V}(\Phi)\Big]+S_{\text{matter}}[e^{2\alpha(\Phi)}g_{\mu\nu},\chi].
	\end{split} \label{action}
	\end{equation}
	This action functional contains six arbitrary functions of one real variable: $\{\mathcal{A},\mathcal{B},\mathcal{C}_1,\mathcal{C}_2,\mathcal{V},\alpha\}$, which after composing with the scalar field $\Phi$ become the scalar functions on the spacetime $M$. They provide, together with the dynamical variables $(\Gamma, g,\Phi)$, the so-called frame for the action (\ref{action}). A change of frame is governed by a consistent action which will be introduced later on. Some of these coefficients have exactly the same meaning as their metric counterparts (c.f. Appendix A), i.e. $\mathcal{A}$ describes coupling between curvature and the field, $\mathcal{B}$ is the kinetic coupling, $\mathcal{V}$ is the potential of self-interaction of the scalar field, while non-zero $\alpha$ means that the action functional features an anomalous coupling between the scalar and matter fields $\chi$. One requires $\mathcal{A}$ be non-negative, otherwise, gravity would be rendered a repulsive force. The coefficients $\mathcal{C}_1$ and $\mathcal{C}_2$ do not have a clear interpretation yet. Their inclusion in the functional is a direct consequence of the Palatini approach we adopted; they do not appear in the metric S-T theory. 
	
	Two vectors $A^\mu_1$ and $A^\mu_2$ are also a novelty. They are constructed purely from metric and linear connection, and their presence is a direct result of lack of a priori established dependence of the connection on the metric tensor. The two vectors are defined to be:
	\begin{subequations}
		\begin{align}
	& A^\mu_1(g,\Gamma)=g^{\mu\nu}g^{\alpha\beta}\nabla_\nu g_{\alpha\beta}=g^{\mu\nu}g^{\alpha\beta}Q_{\nu\alpha\beta}, \\
	& A^\mu_2(g,\Gamma)=-g^{\mu\nu}g^{\alpha\beta}\nabla_\alpha g_{\nu\beta}=-g^{\mu\nu}g^{\alpha\beta}Q_{\alpha\nu\beta}.
		\end{align}
	\end{subequations}
	The $\nabla$ operator is defined with respect to the independent connection, hence covariant derivative of the metric tensor does not have to vanish in general. The extent to which theory fails to be metric is quantified by the so-called non-metricity tensor $Q_{\alpha\mu\nu}=\nabla_\alpha g_{\mu\nu}$. 
	
	The form of the action functional follows necessarily from our requirement that the action remain form-invariant under conformal and almost-geodesic transformations, accompanied by a re-parametrization of the scalar field. This condition states that if one changes the metric tensor, the connection and the scalar field according to the transformation relations given below (we shall call such transformation "changing the frame", and the choice of particular metric, connection and scalar field - "(conformal) frame"), solutions to the field equations are mapped into corresponding solutions obtained in the transformed frame. 
	
	Palatini approach is based on the assumption that the metric and the symmetric connection are independent quantities and thus should transform independently of each other. In the standard approach only the metric tensor is transformed, and the Levi-Civita connection, being a function of the metric, changes accordingly. In our case, one must devise a way to transform these two objects separately, as it should be possible, for instance, to conformally transform the metric while keeping the connection intact. We introduce the following transformations (c.f. \cite{Ulf2}):
	\begin{subequations}
			\begin{align}
			& \bar{g}_{\mu\nu}=e^{2\gamma_1(\Phi)}g_{\mu\nu}, \label{e1} \\ 
			& \bar{\Gamma}^\alpha_{\mu\nu}=\Gamma^\alpha_{\mu\nu}+2 \delta^\alpha_{(\mu}\partial_{\nu)}\gamma_2(\Phi)-g_{\mu\nu}g^{\alpha\beta}\partial_\beta\gamma_3(\Phi), \label{e2}\\
			& \bar{\Phi}=f(\Phi). \label{e3}
			\end{align}
		\end{subequations}
These transformations are invertible:
		\begin{subequations}
			\begin{align}
			& g_{\mu\nu}=e^{2\check{\gamma}_1(\bar{\Phi})}\bar{g}_{\mu\nu}, \label{e12} \\ 
			& \Gamma^\alpha_{\mu\nu}=\bar{\Gamma}^\alpha_{\mu\nu}+2 \delta^\alpha_{(\mu}\partial_{\nu)}\check{\gamma}_2(\bar{\Phi})-\bar{g}_{\mu\nu}\bar{g}^{\alpha\beta}\partial_\beta\check{\gamma}_3(\Phi), \label{e22}\\
			& \Phi=\check{f}(\bar{\Phi}), \label{e32}
			\end{align}
		\end{subequations}
		so that the transformations and their inverse are related in the following way:
		\begin{subequations}
			\begin{align}
			&\check \gamma_i^{}=-\gamma_i \circ f, \label{inverse1}\\
			 & \check{f} =f^{-1}.\label{inverse2}
			\end{align}
		\end{subequations}
The transformations are governed by three smooth functions of the scalar field: $\{\gamma_1,\gamma_2,\gamma_3\}$, depending on the space-time position indirectly, through the scalar field $\gamma_i(\Phi(x))$. Eq.  (\ref{e3})  provides the possibility of field re-definition by the diffeomorphism $f\in  \mathtt{Diff}^{}(\mathbb{R})$ (see Appendix B). Eq. (\ref{e1}) clearly represents the conformal transformation of the metric tensor. It can be further generalized to include the disformal transformations of the metric tensor, given by: $$g_{\mu\nu}=e^{2\check{\gamma}_1(\bar{\Phi})}\bar{g}_{\mu\nu} + D(\bar{\Phi})\partial_\mu \bar{\Phi} \partial_\nu \bar{\Phi},$$ with a disformal factor $D(\bar{\Phi})$; for an example of disformal tranformation use within the Palatini framework, see \cite{dmitri}. In this paper, however, we limit our attention to the case when $D(\bar{\Phi})=0$.  

Eq. (\ref{e2}) is called a generalized almost-geodesic transformation of type $\pi_3$; the word "almost" suggests that one needs to distinguish between the transformation (\ref{e2}) and a transformation which genuinely preserves geodesics on the space-time (see Appendix D). In fact, if the function $\gamma_3$ was equal zero, one would have precisely the geodesic transformation of the affine connection. The new connection preserves also the light cones, leaving the causal structure of spacetime unchanged. If all functions $\gamma_i$ were equal, one would recover standard conformal transformation formulae, identical to the case when the connection is Levi-Civita with respect to the metric tensor. One can also think of the transformation as Weyl transformation, i.e. without assuming that the connection is metric; in particular setting $\gamma_1\neq\gamma_2=\gamma_3$.
	
	One obtains field equations in the standard way, varying with respect to all independent variables entering the action. Unlike in the metric approach, now it is also necessary to vary w.r.t. the linear connection. Three sets of resulting equations are given below:
	\begin{equation}
	\begin{split}
	&\text{\textbf{Metric:}}\\
	& -\frac{1}{2}g_{\mu\nu}\mathcal{L}(\Phi,g,\Gamma)+\mathcal{A}(\Phi)R_{(\mu\nu)}(\Gamma)-\mathcal{B}(\Phi)\partial_\mu\Phi\partial_\nu\Phi\\
	& +\mathcal{C}'_2(\Phi)\partial_\mu\Phi\partial_\nu\Phi-\mathcal{C}'_1(\Phi)g_{\mu\nu}g^{\sigma\beta}\partial_\sigma\Phi\partial_\beta\Phi+\mathcal{C}_2(\Phi)\nabla_\mu\nabla_\nu\Phi-\mathcal{C}_1(\Phi)g_{\mu\nu}\Box \Phi \\
	& +Q_{\beta\lambda\zeta}\partial_\sigma\Phi\Big[\frac{1}{2}\mathcal{C}_2(\Phi)\delta^\sigma_{(\mu}\delta^\beta_{\nu)}g^{\lambda\zeta}-\mathcal{C}_1(\Phi)\left(\frac{1}{2}g_{\mu\nu}g^{\sigma\beta}g^{\lambda\zeta}-g_{\mu\nu}g^{\sigma\lambda}g^{\beta\zeta}+\delta^\sigma_{(\mu}\delta^\beta_{\nu)}g^{\lambda\zeta}\right)\Big]=\kappa^2 T_{\mu\nu},
	\end{split}
	\end{equation}
	
	\begin{equation}
	\begin{split}
	&\text{\textbf{Connection:}}\\
	& \nabla_\alpha\left[\sqrt{-g}\left(g^{\alpha(\zeta}\delta^{\lambda)}_{\beta}-g^{\lambda\zeta}\delta^\alpha_{\beta}\right)\right]=\\
	&=\sqrt{-g}\partial_\alpha\Phi\left[g^{\alpha(\zeta}\delta^{\lambda)}_{\beta}\left(\frac{\mathcal{C}_2(\Phi)-2\mathcal{C}_1(\Phi)-\mathcal{A}'(\Phi)}{\mathcal{A}(\Phi)}\right)-g^{\lambda\zeta}\delta^\alpha_{\beta}\left(\frac{-\mathcal{C}_2(\Phi)-\mathcal{A}'(\Phi)}{\mathcal{A}(\Phi)}\right)\right],
	\end{split}
	\end{equation}
	
	\begin{equation}
	\begin{split}
	&\text{\textbf{Scalar field:}}\\
	& \mathcal{A}'(\Phi)R(g,\Gamma)+\mathcal{B}'(\Phi)g^{\mu\nu}\partial_\mu\Phi\partial_\nu\Phi+2\mathcal{B}(\Phi)\Box\Phi+2\mathcal{B}(\Phi)\partial_\mu\Phi Q_{\nu\alpha\beta}\left(\frac{1}{2}g^{\mu\nu}g^{\alpha\beta}-g^{\alpha\mu}g^{\beta\nu}\right)\\
	& +\frac{1}{\sqrt{-g}}\left[\mathcal{C}_1(\Phi)\nabla_\mu\left(\sqrt{-g}A^\mu_1(g,\Gamma)\right)+\mathcal{C}_2(\Phi)\nabla_\mu\left(\sqrt{-g}A^\mu_2(g,\Gamma)\right)\right]-\mathcal{V}'(\Phi)=2\alpha'(\Phi)T,
	\end{split}
	\end{equation}
	
	\noindent where $T_{\mu\nu}=-\frac{2}{\sqrt{-g}}\frac{\delta(\sqrt{-g}\mathcal{L}_{\text{matter}})}{\delta g^{\mu\nu}}$, $\mathcal{L}$ is simply the gravitational part of Lagrangian; furthermore, all primes denote differentiation with respect to the scalar field $\Phi$. 
	
		An analysis of the equations written above will not be particularly illuminating unless one inspects the equation resulting from varying with respect to the affine connection. As it turns out, it is always possible to find a frame in which the independent connection is the Levi-Civita connection of the metric tensor $g_{\mu\nu}$. One transforms the connection using Eq. (\ref{e2}), with $\check{\gamma}_2$ and $\check{\gamma}_3$ specified by the field equations. Denoting the Levi-Civita connection of the metric tensor $g_{\mu\nu}$ by $\Big\{\genfrac{}{}{0pt}{}{\alpha}{\mu\nu}\Big\}_g$, we find out that it is related to the initial independent affine connection in the following way:
		\begin{equation}
		\begin{split}
		\Gamma^\alpha_{\mu\nu}=\Big\{\genfrac{}{}{0pt}{}{\alpha}{\mu\nu}\Big\}_g+
		\mathcal{F}_1(\Phi)\delta^\alpha_{(\mu}\partial_{\nu)}\Phi-\mathcal{F}_2(\Phi)
		g_{\mu\nu}g^{\alpha\beta}\partial_\beta \Phi ,
		\end{split}\label{sol}
		\end{equation} 
		where the functions $\mathcal{F}_1, \mathcal{F}_2$ of the scalar field $\Phi$ take the form:
		$$\mathcal{F}_1(\Phi)=\frac{2\mathcal{C}_1(\Phi)+(n-3)\mathcal{C}_2(\Phi)+(n-1)\mathcal{A}'(\Phi)}{\mathcal{A}(\Phi)(n-1)(n-2)}$$ and 
		$$\mathcal{F}_2(\Phi)=\frac{2\mathcal{C}_1(\Phi)-\mathcal{C}_2(\Phi)+\mathcal{A}'(\Phi)}{\mathcal{A}(\Phi)(n-2)} .$$ This result simply means that one can always choose a frame in which the theory is effectively metric, with vanishing vectors $A^\mu_1,\:A^\mu_2$. More generally, if $\mathcal{C}_1=\mathcal{C}_2\equiv\mathcal{C}$, then one has $\mathcal{F}_1=\mathcal{F}_2\equiv\mathcal{F}=\frac{\mathcal{C}(\Phi)+\mathcal{A}'(\Phi)}{\mathcal{A}(\Phi)(n-2)}$ and the metric providing the connection has the form $\exp\left(\int \mathcal{F}(\Phi)d\Phi\,\right) g_{\mu\nu}$.  This gives a link to the so-called C-theories of gravity studied recently in \cite{K1}-\cite{K3}.
	
	Since the connection can be always solved in terms of the metric and the scalar field, there are no additional physical degrees of freedom carried by it. The connection always turns out to be an auxiliary field \cite{olmo2009}.	
	
	The relation (\ref{sol}) is defined by two functions, which in general (except the case mentioned above) are not equal. One can identify them as the functions $\check{\gamma}_2$ and $\check{\gamma}_3$ relating affine connections of two different frames. Frame, in which the theory turns out to be fully metric, can be obtained by plugging back the connection (\ref{sol}) in the action functional (\ref{action}). Such a change of frame should not affect the form of action functional (otherwise solutions of equations of motion in one frame would not be mapped to solution in another frame, which would contradict one of our basic assumptions), and the coefficients $\{\mathcal{A},\mathcal{B},\mathcal{C}_1,\mathcal{C}_2,\mathcal{V},\alpha\}$ will change in a way that preserves the functional form of the action. Exact transformation relations will be presented in the next section.
	
	Because the transformation (\ref{e2}) depends on two independent parameters, one cannot in general end up in a frame in which the initial independent connection is Levi-Civita with respect to \textit{some} metric tensor, as the transformation of the metric is governed by a single function $\check{\gamma}_1$. However, if $\mathcal{C}_1=\mathcal{C}_2$, then it is possible to transform the metric tensor in such a way that the initial independent connection becomes a Levi-Civita connection of the transformed, new metric.

\section{Transformation formulae}
	
 	Redefinition of the transformations leads to a modification of conformal mapping formulae for all quantities built from the connection, i.e. Riemann tensor and its contractions. This is an obvious consequence of decoupling metric tensor from the connection. In the metric approach, transformation of the Riemann tensor is fully determined by the way the metric transforms; here, one must take into account the fact that the transformation is governed by the functions $\check{\gamma}_2$ and $\check{\gamma}_3$. Additionally, covariant derivative of the metric does not vanish in general, and this fact plays an important role in the process of deriving transformation relations. If the calculations are performed in $n$ dimensions, requiring the transformations be defined by Eq. (\ref{e1})-(\ref{e3}), the formulae relating Riemann tensors of two different conformal frames are the following:
	\begin{equation}
	\begin{split}
	R^\alpha_{\mu\beta\nu}& =\bar{R}^\alpha_{\mu\beta\nu}+\delta^\alpha_\nu\bar{\nabla}_\beta\bar{\nabla}_\mu\check{\gamma}_2(\bar{\Phi})-\delta^\alpha_\beta\bar{\nabla}_\nu\bar{\nabla}_\mu\check{\gamma}_2(\bar{\Phi})-\delta^\alpha_\nu\bar{\nabla}_\beta\check{\gamma}_2(\bar{\Phi})\bar{\nabla}_\mu\check{\gamma}_2(\bar{\Phi})+\delta^\alpha_\beta\bar{\nabla}_\nu\check{\gamma}_2(\bar{\Phi})\bar{\nabla}_\mu\check{\gamma}_2(\bar{\Phi})\\
	& +\bar{g}_{\mu\beta}\bar{g}^{\alpha\lambda}\bar{\nabla}_\nu\bar{\nabla}_\lambda\check{\gamma}_3(\bar{\Phi})-\bar{g}_{\mu\nu}\bar{g}^{\alpha\lambda}\bar{\nabla}_\beta\bar{\nabla}_\lambda\check{\gamma}_3(\bar{\Phi})+\delta^\alpha_\nu\bar{g}_{\mu\beta}\bar{g}^{\sigma\lambda}\bar{\nabla}_\sigma\check{\gamma}_3(\bar{\Phi})\bar{\nabla}_\lambda\check{\gamma}_2(\bar{\Phi})\\
	&-\delta^\alpha_\beta\bar{g}_{\mu\nu}\bar{g}^{\sigma\lambda}\bar{\nabla}_\sigma\check{\gamma}_3(\bar{\Phi})\bar{\nabla}_\lambda\check{\gamma}_2(\bar{\Phi}) +  \bar{g}^{\alpha\lambda}\bar{g}_{\mu\nu}\bar{\nabla}_\lambda\bar{\gamma}_3(\bar{\Phi})\bar{\nabla}_\beta\check{\gamma}_3(\bar{\Phi})-\bar{g}^{\alpha\lambda}\bar{g}_{\mu\beta}\bar{\nabla}_\lambda\check{\gamma}_3(\bar{\Phi})\bar{\nabla}_\nu\check{\gamma}_3(\bar{\Phi}) \\
	&+ \bar{g}^{\alpha\lambda}\bar{\nabla}_\nu\bar{g}_{\mu\beta}\bar{\nabla}_\lambda\check{\gamma}_3(\bar{\Phi})-\bar{g}^{\alpha\lambda}\bar{\nabla}_\beta\bar{g}_{\mu\nu	}\bar{\nabla}_\lambda\check{\gamma}_3(\bar{\Phi})+\bar{g}_{\mu\beta}\bar{\nabla}_\nu\bar{g}^{\alpha\lambda}\bar{\nabla}_\lambda\check{\gamma}_3(\bar{\Phi})-\bar{g}_{\mu\nu}\bar{\nabla}_\beta\bar{g}^{\alpha\lambda}\bar{\nabla}_\lambda\check{\gamma}_3(\bar{\Phi}).
	\end{split}
	\end{equation}
	
	The formula for the (symmetrized) Ricci curvature tensor reads as follows:
	
	\begin{equation}
	\begin{split}
	R_{(\mu\nu)} & =\bar{R}_{(\mu\nu)}-(n-1)\bar{\nabla}_{\mu}\bar{\nabla}_\nu\check{\gamma}_2(\bar{\Phi})+\bar{\nabla}_{\mu}\bar{\nabla}_\nu\check{\gamma}_3(\bar{\Phi})+(n-1)\bar{\nabla}_\nu\check{\gamma}_2(\bar{\Phi})\bar{\nabla}_\mu\check{\gamma}_2(\bar{\Phi})\\
	& -\bar{\nabla}_\nu\check{\gamma}_3(\bar{\Phi})\bar{\nabla}_\mu\check{\gamma}_3(\bar{\Phi})-\bar{g}_{\mu\nu}\bar{g}^{\alpha\beta}\bar{\nabla}_{\alpha}\bar{\nabla}_\beta\check{\gamma}_3(\bar{\Phi})-(n-1)\bar{g}_{\mu\nu}\bar{g}^{\alpha\beta}\bar{\nabla}_\alpha\check{\gamma}_3(\bar{\Phi}){\nabla}_\beta\check{\gamma}_2(\bar{\Phi})\\
	 &+\bar{g}_{\mu\nu}\bar{g}^{\alpha\beta}\bar{\nabla}_\alpha\check{\gamma}_3(\bar{\Phi}){\nabla}_\beta\check{\gamma}_3(\bar{\Phi})  +\Big[\bar{g}_{\mu\nu}\bar{g}^{\alpha\beta}\bar{g}^{\sigma\lambda}\bar{\nabla}_\alpha\bar{g}_{\beta\sigma}-\bar{g}^{\alpha\lambda}\bar{\nabla}_\alpha\bar{g}_{\mu\nu}\Big]\bar{\nabla}_\lambda\check{\gamma}_3(\bar{\Phi}).
	\end{split}
	\end{equation}	
%
Finally, contracting the previous formula with the metric tensor, we get an expression for the Palatini-Ricci scalar:
	 \begin{equation}
	\begin{split}
	R& =e^{-2\check{\gamma}_1(\bar{\Phi})}\Big[\bar{R}-(n-1)\bar{g}^{\mu\nu}\bar{\nabla}_\mu\bar{\nabla}_\nu\left(\check{\gamma}_2(\bar{\Phi})+\check{\gamma}_3(\bar{\Phi})\right)+\bar{g}^{\mu\nu}\bar{g}^{\lambda\sigma}\Big(n\bar{\nabla}_\mu\bar{g}_{\nu\sigma}-\bar{\nabla}_\sigma\bar{g}_{\nu\mu}\Big)\bar{\nabla}_\lambda\check{\gamma}_3(\bar{\Phi}) \\
	& +(n-1)\bar{g}^{\mu\nu}\left(\bar{\nabla}_\mu\check{\gamma}_2(\bar{\Phi})\bar{\nabla}_\nu\check{\gamma}_2(\bar{\Phi}) - n\bar{\nabla}_\mu\check{\gamma}_2(\bar{\Phi})\bar{\nabla}_\nu\check{\gamma}_3(\bar{\Phi})+\bar{\nabla}_\mu\check{\gamma}_3(\bar{\Phi})\bar{\nabla}_\nu\check{\gamma}_3(\bar{\Phi})\right)\Big]. \label{curvsc}
	\end{split}
	\end{equation}
In the Weyl case $\gamma_3=\gamma_2+\text{const}$ one gets
\begin{equation}
	\begin{split}
	R& =e^{-2\check{\gamma}_1(\bar{\Phi})}\Big[\bar{R}-2(n-1)\bar{g}^{\mu\nu}\bar{\nabla}_\mu\bar{\nabla}_\nu\check{\gamma}_2(\bar{\Phi})+\bar{g}^{\mu\nu}\bar{g}^{\lambda\sigma}\Big(n\bar{\nabla}_\mu\bar{g}_{\nu\sigma}-\bar{\nabla}_\sigma\bar{g}_{\nu\mu}\Big)\bar{\nabla}_\lambda\check{\gamma}_2(\bar{\Phi}) \\
	& -(n-1)(n-2)\bar{g}^{\mu\nu}\bar{\nabla}_\mu\check{\gamma}_2(\bar{\Phi})\bar{\nabla}_\nu\check{\gamma}_2(\bar{\Phi})\Big]. \label{curvsc_2}
	\end{split}
	\end{equation}
	When $\gamma_2+\gamma_3=\text{const}$ the expression (\ref{curvsc}) reduces instead to
	\begin{equation}
	\begin{split}
	R& =e^{-2\check{\gamma}_1(\bar{\Phi})}\Big[\bar{R}+\bar{g}^{\mu\nu}\bar{g}^{\lambda\sigma}\Big(n\bar{\nabla}_\mu\bar{g}_{\nu\sigma}-\bar{\nabla}_\sigma\bar{g}_{\nu\mu}\Big)\bar{\nabla}_\lambda\check{\gamma}_2(\bar{\Phi}) \\
	& +(n-1)(n+2)\bar{g}^{\mu\nu}\bar{\nabla}_\mu\check{\gamma}_2(\bar{\Phi})\bar{\nabla}_\nu\check{\gamma}_2(\bar{\Phi})\Big]. \label{curvsc_3}
	\end{split}
	\end{equation}

Since the functions $\check{\gamma}_2$ and $\check{\gamma}_3$ do not depend on the spacetime position explicitly, derivatives of these quantities can be cast in the following form: 
	$$\bar{\nabla}_\mu\check{\gamma}_i(\bar{\Phi})=\frac{d \check{\gamma}_i(\bar{\Phi})}{d\bar{\Phi}}\bar{\nabla}_\mu\bar{\Phi}\equiv\check{\gamma}'_i\bar{\nabla}_\mu\bar{\Phi},$$
	where $i=2,3$. 
	 
	Conformal transformation and almost-geodesic mapping, accompanied by re-definition of the scalar field, applied to the three independent variables should map solutions of equations of motion in one frame to corresponding solutions in another frame. For it to be true, the way functions $\{\mathcal{A},\ldots,\alpha\}$ transform must be governed by equations analogous to (\ref{eqns7}), as the action functional needs to preserve its form. The condition of form-invariance of the action leads to the following transformation equations for the five independent scalar field functions:
	\begin{subequations}
		\begin{align}
		\mathcal{\bar{A}}(\bar{\Phi})&=e^{(n-2)\check{\gamma}_1(\bar{\Phi})}\mathcal{A}(\check{f}(\bar{\Phi})), \label{t1}\\
			\begin{split}
			\mathcal{\bar{B}}(\bar{\Phi})&=e^{(n-2)\check{\gamma}_1(\bar{\Phi})}\Big[\mathcal{B}(\check{f}(\bar{\Phi}))(\check{f}'(\bar{\Phi}))^2+(n-1)\Big(n\mathcal{A}(\check{f}(\bar{\Phi}))\check{\gamma}'_2(\bar{\Phi})\check{\gamma}'_3(\bar{\Phi})-\mathcal{A}(\check{f}(\bar{\Phi}))\left(\check{\gamma}'_2(\bar{\Phi})\right)^2\\
			&-\mathcal{A}(\check{f}(\bar{\Phi}))\left(\check{\gamma}'_3(\bar{\Phi})\right)^2-\frac{d\mathcal{A}(\check{f}(\bar{\Phi}))}{d\bar{\Phi}}(\check{\gamma}'_2(\bar{\Phi})+\check{\gamma}'_3(\bar{\Phi}))\\
			& -(n-2)\mathcal{A}(\check{f}(\bar{\Phi}))\check{\gamma}'_1(\bar{\Phi})(\check{\gamma}'_2(\bar{\Phi})+\check{\gamma}'_3(\bar{\Phi}))\Big)\\
			 &+\check{f}'(\bar{\Phi})\Big(\mathcal{C}_1(\check{f}(\bar{\Phi}))(2 n\check{\gamma}'_1(\bar{\Phi})-2(n+1)\check{\gamma}'_2(\bar{\Phi})+2\check{\gamma}'_3(\bar{\Phi}))\\
			 &-\mathcal{C}_2(\check{f}(\bar{\Phi}))(2\check{\gamma}'_1(\bar{\Phi})-(n+3)\check{\gamma}'_2(\bar{\Phi})+(n+1)\check{\gamma}'_3(\bar{\Phi}))\Big)\Big],
			\end{split}
\end{align}
\begin{align}
		\mathcal{\bar{C}}_1(\bar{\Phi})&=e^{(n-2)\check{\gamma}_1(\bar{\Phi})}\Big[\check{f}'(\bar{\Phi})\mathcal{C}_1(\check{f}(\bar{\Phi}))-\mathcal{A}(\check{f}(\bar{\Phi}))\left(\frac{n-1}{2}\check{\gamma}'_2(\bar{\Phi})+\frac{n-3}{2}\check{\gamma}'_3(\bar{\Phi})\right)\Big],\\
		\mathcal{\bar{C}}_2(\bar{\Phi})&=e^{(n-2)\check{\gamma}_1(\bar{\Phi})}\Big[\check{f}'(\bar{\Phi})\mathcal{C}_2(\check{f}(\bar{\Phi}))-\mathcal{A}(\check{f}(\bar{\Phi}))\left((n-1)\check{\gamma}'_2(\bar{\Phi})-\check{\gamma}'_3(\bar{\Phi})\right)\Big],\\
		\mathcal{\bar{V}}(\bar{\Phi})&=e^{n\check{\gamma}_1(\bar{\Phi})}\mathcal{V}(\check{f}(\bar{\Phi})), \\
		\bar{\alpha}(\bar{\Phi})&=\alpha(\check{f}(\bar{\Phi}))+\check{\gamma}_1(\bar{\Phi}).\label{t6}
		\end{align} \label{transformations}
	\end{subequations}
	These transformations are induced by the transformations (\ref{e1})-(\ref{e3}) of independent variables which are invertible. This means that (\ref{t1})-(\ref{t6}) allow us to transform solutions obtained in one frame into another, therefore we have split theories given by the action (\ref{action})  into classes which are solution-equivalent. Next task is to find a typical representative in each class. One choice mentioned before is the so-called Einstein frame, another one is known as the Jordan frame.
	
	As we can see, some of the transformation relations involve nothing but a simple multiplication of the "old" coefficients by a factor related to the transformation of the metric tensor. These relations do not depend on the approach we adopt - they retain the same form regardless of whether we work within metric or Palatini formalism. However, coefficients $\mathcal{C}_1, \mathcal{C}_2$ and $\mathcal{B}$ transform in a more complicated way depending on whether the theory is metric or not. The transformation relations preserve the sign of the $\mathcal{A}$ coefficient. Similarly, if $\mathcal{B}$ is subject to a scalar field re-parametrization only, then its sign does not change as well. By the same token, if the potential $\mathcal{V}$ vanishes in one frame, it cannot emerge in any other.
	
	Due to our freedom of choice of three functions $\{\gamma_1,\gamma_2,\gamma_3\}$ and re-parametrization of the scalar field $\Phi=\check{f}(\bar{\Phi})$, it is always possible to fix four of the above six coefficients. We shall call such fixing "choosing a frame", as it was mentioned before. If we specify the remaining two functions, we choose a theory. For example, the four functions $\{\gamma_1,\gamma_2,\gamma_3,f\}$ can be chosen in such a way that four coefficients $\{\mathcal{B},\mathcal{C}_1,\mathcal{C}_2,\alpha\}$ vanish, simplifying the calculations. Results obtained in a given frame can be always "translated" to another frame if the two frames can be related by a conformal transformation accompanied by a re-parametrization of the scalar field. It must be also noted that increased number of functions used to change the frame (from two in scalar-tensor theory in the metric approach - see Appendix A - to four in case of the Palatini formalism) result in additional coefficients appearing in the action functional. However, analogously to the metric case, despite the fact we are able to fix four of them, we are always left with two functions, defining the particular theory. 
	
	Conformal and generalized almost-geodesic transformation establish a mathematical equivalence of two frames. On the physical ground, they may constitute two very different theories. The multitude of equivalent theories poses a problem of identifying frames which can be related by the transformations given by Eqs (\ref{e1})-(\ref{e3}). Such frames may bear no resemblance to one another and yet, be two different manifestations of the same theory, but written using different variables. This situation suggests that it would be desirable to formulate the general scalar-tensor theory in a frame-independent way, fully analogous to the way GR circumvents the problem of deciding upon the "right" coordinate system to describe physical phenomena by resorting to the language of tensors, allowing one to write equations in a covariant manner. In case of scalar-tensor gravity in the Palatini approach, we decided to follow on \cite{jarv} and find invariant quantities built from coefficients $\{\mathcal{A},\ldots,\alpha\}$, metric and connection, whose values are independent of the choice of frame - just like, for instance, value of $R^\alpha_{\:\:\mu\beta\nu}R_{\alpha}^{\:\:\mu\beta\nu}$ does not depend on our choice of coordinate frame. This analogy, however, should not be taken too seriously, as general covariance in case of GR is a consequence of the fact that our description of Nature should not depend on an artificial construct of coordinate frame, whereas such invariance of physical laws is not present when changing conformal frames. For example, geodesic curves, due to covariant formulation of geodesic equations, are the same in every coordinate frame; on the other hand, if the mapping (\ref{e2}) is applied, geodesics are not preserved (unless $\gamma_3=0$), thus leading to emergence of an unobserved "fifth force", causing particles to deviate from their standard trajectories, see e.g. \cite{SWB18} for application to explaining galaxy rotational curves.

	\section{Invariant quantities and their applications}
	 In order to check whether two frames can be conformally related, we may introduce the notion of invariants \cite{jarv}. The invariants are quantities which are built from the functions $\{\mathcal{A},\mathcal{B},\mathcal{C}_1,\mathcal{C}_2,\mathcal{V},\alpha\}$ such that their functional dependence on them is the same in every frame. Also, their value at a given spacetime point remains unchanged. If the invariants calculated for one theory coincide with the invariant quantities computed for another one, we can always find a conformal transformation relating these two theories (this transformation, however, may not obey group composition law, and the solutions to equations in both frames may not be mathematically equivalent). The way the invariants are constructed comes from transformation properties of the five arbitrary functions. Some of the functions get multiplied only by a factor, while the coefficients $\mathcal{B}$, $\mathcal{C}_1$ and $\mathcal{C}_1$ transform in a more sophisticated manner. Taking this into account, we can find the correct combinations of the functions giving us quantities expressed in terms of the same coefficients irrespective of the frame we are in. Two exemplary invariants are given below\footnote{In \cite{jarv}, this invariant is defined as $\mathcal{I}_1(\Phi)=\frac{e^{2\alpha(\Phi)}}{\mathcal{A}(\Phi)}$ (in four dimensions).}:
	 
	 \begin{equation}
	 \mathcal{I}_1(\Phi)=\frac{\mathcal{A}(\Phi)}{e^{(n-2)\alpha(\Phi)}}, \label{i1}
	 \end{equation}
	 \begin{equation}
	 \mathcal{I}_2(\Phi)=\frac{\mathcal{V}(\Phi)}{(\mathcal{A}(\Phi))^{\frac{n}{n-2}}}. \label{i2}
	 \end{equation}
	 
	 In four dimensions, the invariant $\mathcal{I}_1$ characterizes the non-minimal coupling \cite{inin}. Apart from the case when $\mathcal{A}=e^{2\alpha}$, its constancy means that both $\mathcal{A}$ and $e^{2\alpha}$ are some numbers, implying that in such theory scalar field is entirely decoupled from curvature and matter. The invariant $\mathcal{I}_2$ generalizes the notion of self-interaction potential.  
	  It should be obvious that any function of the invariants is invariant itself. Moreover, spacetime derivatives of the invariants are invariant, as well as derivatives with respect to other invariants (if we treat an invariant as a function of another invariant quantity) \cite{jarv}. 
	 It is also possible to construct invariant metrics and connections. In the case of the metric there is no unique way of doing so, but in this paper, only two possibilities will be considered:
	 \begin{equation}
	 \hat{g}_{\mu\nu}=(\mathcal{A}(\Phi))^{\frac{2}{n-2}}g_{\mu\nu}, \label{g1}
	 \end{equation}
	 or
	 \begin{equation}
	 \tilde{g}_{\mu\nu}=e^{2\alpha(\Phi)}g_{\mu\nu}. \label{g2}
	 \end{equation}	As for the affine connection, it is possible to choose the following:
	 \begin{equation}
	 \hat{\Gamma}^\alpha_{\mu\nu}=\Gamma^\alpha_{\mu\nu}-2\mathcal{P}_1(\Phi)\delta^\alpha_{(\mu}\partial_{\nu)}\Phi +g_{\mu\nu}g^{\alpha\beta}\mathcal{P}_2(\Phi)\partial_\beta\Phi\,, \label{con}
	 \end{equation}
	 where:
	 $$\mathcal{P}_1(\Phi)=\frac{2 \mathcal{C}_1(\Phi) + (n-3)\mathcal{C}_2(\Phi)}{ \mathcal{A}(\Phi)(n-1)(n-2)}$$
	 and
	 $$\mathcal{P}_2(\Phi)=\frac{-2 \mathcal{C}_1(\Phi) +\mathcal{C}_2(\Phi)}{ \mathcal{A}(\Phi)(n-2 )}\,.$$
	 From a purely algebraic point of view, invariance of the quantities given above means that when changing the frame, the additional terms multiplying the metric or added to the connection transform in a way balancing out multiplicative or additive terms containing transformation-defining functions $\check{\gamma}_1$, $\check{\gamma}_2$ and $\check{\gamma}_3$. Their physical invariance is much more profound a can be a subject for various phenomenological speculations (see e.g. \cite{EPS}-\cite{LF2}). It is obvious that conformal transformation of the metric tensor does not preserve the line element on a (pseudo-)Riemannian manifold due to the fact that conformal change is not equivalent to a simple coordinate transformation. Thence, two observers using conformally-related metric tensors will agree only on the causal structure of space-time but will measure distances differently; the same can be said about affine connections used to determine geodesic curves. Observers of different frames will, in general, disagree on whether a test particle moves along its geodesic, as the general almost-geodesic mapping (or conformal transformation in case of the purely metric approach) changes geodesics (except for the null ones) on a given space-time. Introduction of invariant metric tensors and connections aims at resolving - at least partially - this ambiguity. If two observers of different frames agree on using the same invariant quantity to describe geometry, the measurements they make shall give exactly the same outcome. In case of the invariant metric, all distances will be the same, while the invariant connection guarantees invariance of geodesic curves. There is, however, more than one invariant metric (and in fact, there are also multiple invariant connections, but in this paper, we introduce only one), so that no unique way of choosing invariant objects to describe the geometry of space-time exists.

	\subsection{Integral invariants}

	Let us define the following quantity \footnote{This is integral invariant, which  is determined up to arbitrary integration constant.
		The choice of the sign  $\pm$ in (\ref{i3}) has to ensure positivity of the expression inside the square root. }:
	\begin{equation}
	\begin{split}
	\mathcal{I}^n_E(\Phi)&=\int\Bigg(\pm\frac{(n-2)\mathcal{A}(\Phi) \mathcal{B}(\Phi) +
		2\mathcal{A}'(\Phi)[
		\mathcal{C}_2(\Phi)  - n\mathcal{C}_1(\Phi)] }{ (n-2)\mathcal{A}(\Phi)^2} \pm\\
	&
	\pm\frac{ (n^2-5) \mathcal{C}_2(\Phi)^2  - 4 \mathcal{C}_1(\Phi)^2+2(4 +  n - n^2)\mathcal{C}_1(\Phi)\mathcal{C}_2(\Phi)\big)}{(n-2) (n-1) \mathcal{A}(\Phi)^2}\Bigg)^{\frac{1}{2}}d\Phi.
	\end{split} \label{i3}
	\end{equation}
	 Such quantity is a genuine invariant for arbitrary transformation $\{f, \gamma_1,\gamma_2,\gamma_3\}
\in\mathtt{Diff}^{(3)}(\mathbb{R})$.
	
	 	In four dimensions, the quantity $\mathcal{I}_E$ \footnote{From now on, all invariants shall be written without the superscript denoting the number of dimensions if $n=4$} can be written as:
	 	
	 \begin{equation}
	 \begin{split}
	 \mathcal{I}_E(\Phi)= & \int\Bigg(\pm\frac{\mathcal{A}(\Phi) \mathcal{B}(\Phi) -\frac{2}{3}\mathcal{C}_1(
	 		\Phi)^2-\frac{8}{3}\mathcal{C}_1(\Phi)\mathcal{C}_2(\Phi)+\frac{11}{6}\mathcal{C}_2(
	 		\Phi)^2 -4 \mathcal{C}_1(\Phi) \mathcal{A}'(\Phi)}{\mathcal{A}(\Phi)^2}\\
	 		& \pm\frac{\mathcal{C}_2(\Phi) \mathcal{A}'(\Phi)}{\mathcal{A}(\Phi)^2}\Bigg)^\frac{1}{2}d\Phi.
	 		\end{split}
	 \end{equation}
It will be shown later on that in the Einstein-like frame it plays the role of the scalar field.
	 
	  In can be noticed that the function $\mathcal{A}(\Phi)$ in the denominator of (\ref{i3}) can be replaced by $e^{(n-2)\alpha(\Phi)}$ without changing its transformation properties. We will arrive at an invariant closely related to $\mathcal{I}^n_E$. Its importance shall be revealed while investigating different frame parametrizations of the S-T theories.
	\begin{equation}
	\begin{split}
	\mathcal{I}^n_J(\Phi)&=\int e^{-\frac{n-2}{2}\alpha(\Phi)}\Bigg(\pm\frac{(n-2)\mathcal{A}(\Phi) \mathcal{B}(\Phi) +
		2\mathcal{A}'(\Phi)[
		\mathcal{C}_2(\Phi)  - n\mathcal{C}_1(\Phi)] }{ (n-2)\mathcal{A}(\Phi)} \pm\\
	&
	\pm\frac{ (n^2-5) \mathcal{C}_2(\Phi)^2  - 4 \mathcal{C}_1(\Phi)^2+2(4 +  n - n^2)\mathcal{C}_1(\Phi)\mathcal{C}_2(\Phi)\big)}{(n-2) (n-1) \mathcal{A}(\Phi)}\Bigg)^{\frac{1}{2}}d\Phi.
	\end{split} \label{i4}
	\end{equation}

	 This invariant was given the subscript "J" to indicate that it arises naturally in the Jordan frame. It is obvious that if $\mathcal{I}^n_E$ vanishes, so does $\mathcal{I}^n_J$. 

	 \section{Einstein and Jordan frames, and their invariant generalizations}
	 So far, we have been using terms "Jordan/Einstein frame" without defining it in an unambiguous way. As it is widely known, the notion of a (conformal) frame has been applied to an analysis of the S-T theories primarily in the metric approach. It is straightforward to extend the concepts of Einstein and Jordan frames to Palatini theory as well. We define the former in the following way:
	\begin{definition}
		The \textbf{Einstein frame in the Palatini theory} is characterized by specific values of four out of six arbitrary functions $\{\mathcal{A},\ldots,\alpha\}$:
		$\mathcal{A}=1,\:\mathcal{B}=\epsilon_\text{Palatini},\:\mathcal{C}_1=\mathcal{C}_2=0.$\\	The action functional is given by:
		\begin{center}
		$
		S[g_{\mu\nu}^E,(\Gamma^E)^\alpha_{\mu\nu},\Phi]=\frac{1}{2\kappa^2}\int_{\Omega}d^nx \sqrt{-g^E}\Big(R(g^E,\Gamma^E)-\epsilon_\text{Palatini} (g^E)^{\mu\nu}\nabla_\mu\Phi\nabla_\nu\Phi-\mathcal{V}(\Phi)\Big)$\\$+S_{\text{matter}}\left[e^{2\alpha(\Phi)}g_{\mu\nu}^E,\chi\right],
		$
		\end{center}
where $\epsilon_\text{Palatini}\equiv (\pm 1,0)$ is a three valued function.  
		\label{definition}\end{definition}
It follows from the very definition that there are three types of Einstein frames, depending on the value of the parameter $\epsilon_\text{Palatini}$,  which cannot transform each other by a diffeomorphism	 \footnote{ However, it can be changed by making use of disformal transformations \cite{dmitri}.}.
In the simplest case $\gamma_1=\gamma_2=\gamma_3=0$ its values  can  be identified with the signature of $\cb$, i.e. $\epsilon_\text{Platini}=\text{sign}(\cb)$.
In fact, Einstein frames can be labelled as a triple $(\epsilon_\text{Palatini}, \cv,\alpha)$. They include the original Einstein-Hilbert-Palatini  action as a particular case: $\epsilon_\text{Palatini}=\cv=\alpha=0$.  One should notice that the frames with  $\epsilon_\text{Palatini}=0$ are singular in the following sense: 
scalar field re-definition by an arbitrary diffeomorphism  $f\in\mathtt{Diff}^{}(\mathbb{R})$
transforms one Einstein frame into another (within the same orbit) without changing the value of $\epsilon_\text{Palatini}=0$. This  is not the case for
$\epsilon_\text{Palatini}=\pm 1$: such frames are not preserved under diffeomorphisms. In the Einstein frame, the choice $\epsilon_\text{Palatini}=+1$ suggests that the scalar field has positive energy, whereas for $\epsilon_\text{Palatini}=-1$, the theory features a ghost \footnote{In the metric case, when one considers weak-field approximation, due to the presence of non-minial coupling, the negative value of the parameter $\epsilon_\text{Palatini}$ does not necessarily mean that the physical, interacting field is a ghost, even if the the inital field $\Phi$ is \cite{fuj}.} \cite{fuj}.

Because the transformations  (\ref{e1})-(\ref{e2}) act in a self-consistent way, any theory has a mathematically equivalent Einstein frame representation. Therefore, all possible scalar-tensor theories in the Palatini approach can be also labelled by the triple $(\epsilon_\text{Palatini}, \mathcal{V}, \alpha)$ in the Einstein frame.

More generally,  one can show (c.f. (\ref{fe3})) that the theory written in the Einstein frame becomes effectively metric. 
	
	 For completeness, let us also write the invariants we have introduced so far for the Einstein frame:
	 \begin{subequations}
	 	\begin{align}
	 	& \mathcal{I}^n_1(\Phi)=e^{-(n-2)\alpha(\Phi)}, \\
	 	& \mathcal{I}^n_2(\Phi)=\mathcal{V}(\Phi), \\
	 	& \mathcal{I}^n_E(\Phi)=\sqrt{\pm \epsilon_\text{Palatini}}(\Phi-\Phi_0).
	 	\end{align}
	\end{subequations}
	As one can see, the quantity $\mathcal{I}^n_E$ plays the role of the scalar field in the Einstein frame.  
	
	In order to understand better how the invariants can be used to find out whether a given theory is equivalent to some other theory written in the Einstein frame via transformations (\ref{e1})-(\ref{e3}), let us consider the following example: an S-T theory is described by the action functional:
	\begin{equation}
	\begin{split}
	 S[\bar{g}_{\mu\nu},\bar{\Gamma}^\alpha_{\mu\nu},\bar{\Phi}]=\frac{1}{2\kappa^2}&\int_{\Omega}d^nx\sqrt{-\bar{g}}\Big[\bar{\mathcal{A}}(\bar{\Phi})R(\bar{g},\bar{\Gamma})-\bar{\mathcal{B}}(\bar{\Phi})\bar{g}^{\mu\nu}\bar{\nabla}_\mu\bar{\Phi}\bar{\nabla}_\nu\bar{\Phi}-\bar{A}_1^\mu(\bar{g},\bar{\Gamma})\bar{\mathcal{C}}_1(\bar{\Phi})\bar{\nabla}_\mu\bar{\Phi}\\
	& -\bar{A}_2^\mu(\bar{g},\bar{\Gamma})\bar{\mathcal{C}}_2(\bar{\Phi})\bar{\nabla}_\mu\bar{\Phi}-\bar{\mathcal{V}}(\bar{\Phi})\Big]+S_{\text{matter}}[e^{2\bar{\alpha}(\bar{\Phi})}\bar{g}_{\mu\nu},\chi].
	\end{split} \label{example}
	\end{equation}
	Such theory always possesses the Einstein frame representation. The comparison of the quantities $\mathcal{I}^n_1$ and $\mathcal{I}^n_2$ will yield the exact form of the $\mathcal{V}$ and $\alpha$ functions in the transformed frame:
	$$\alpha(\Phi)=\bar{\alpha}(\bar{\Phi}(\Phi))-\frac{1}{n-2}\ln \bar{\mathcal{A}}(\bar{\Phi}(\Phi)),$$
	$$\mathcal{V}(\Phi)=\frac{\bar{\mathcal{V}}(\bar{\Phi}(\Phi))}{\big(\bar{\mathcal{A}}(\bar{\Phi}(\Phi))\big)^{\frac{n}{n-2}}},$$
	where $\Phi$ is the scalar field in the new frame; it becomes a function of the "old" scalar field $\bar{\Phi}$. 

	The Jordan frame is defined as follows:
		\begin{definition}
	The \textbf{Jordan frame in the Palatini theory} is characterized by specific values of four out of the six arbitrary functions $\{\mathcal{A},\ldots,\alpha\}$:
			$\mathcal{A}=\Psi,\:\mathcal{C}_1=\mathcal{C}_2=\alpha=0$.\\
			The action functional is given by:
			\begin{center}
			$
			S[g_{\mu\nu}^J,(\Gamma^J)^\alpha_{\mu\nu},\Psi]=\frac{1}{2\kappa^2}\int_{\Omega}d^nx \sqrt{-g^J}\Big(\Psi R(g^J,\Gamma^J)-\mathcal{B}(\Psi)(g^J)^{\mu\nu}\nabla_\mu\Psi\nabla_\nu\Psi-\mathcal{U}(\Psi)\Big)$ $+S_{\text{matter}}\left[g_{\mu\nu}^J,\chi\right].
			$
			\end{center}
		\end{definition}
Therefore, the Jordan frame can be described by 		two functions $(\cb, \mathcal{U})$.  
In the Jordan frame, there is no coupling between the scalar field and matter; the field - or a function of it, but it can always be re-defined appropriately - is coupled directly to the curvature. We impose no conditions on the kinetic coupling $\mathcal{B}$ and the potential $\mathcal{U}$. It can be shown, varying the action expressed in the Jordan frame w.r.t. all dynamical variables, that the curvature scalar is in fact built from a metric conformally related to the initial one. Thence, the Jordan frame in the Palatini approach is in fact almost identical to its metric counterpart, except for a difference in the kinetic coupling. This difference is simply a Brans-Dicke term $\frac{\omega}{\Psi}$, where $\omega$ is a constant and depends on the number of dimensions. This term shall be given explicitly later on when considering the invariant generalizations of the Jordan frame.

	 We may now attempt  
 to express the action (\ref{action}) for S-T theories fully in terms of invariant quantities. Such an approach would be advantageous because any computations performed in an invariant - or generalized - frame will become independent of the variables we use. Unfortunately, there is no unique way of choosing an invariant frame, as one needs to choose between two invariant metric tensors that have been introduced. The existence of (at least) two non-equivalent invariant metric tensors forces us to analyze the theory in two distinct invariant  frames. In each frame, we shall be using the invariant connection $\hat{\Gamma}$ given by (\ref{con}). If we decide to use the variables $(\hat{g},\hat{\Gamma},\mathcal{I}^{n}_E)$ (assuming that the relation (\ref{i3}) between the invariant $\mathcal{I}^n_E$ and the scalar field $\Phi$ is invertible; see \cite{jarv}), the action functional (\ref{action}) will take on the following Einstein frame form:
	 \begin{equation}
	 S[\hat{g}_{\mu\nu},\hat{\Gamma}^\alpha_{\mu\nu},\mathcal{I}^n_E]=\frac{1}{2\kappa^2}\int_{\Omega}d^nx\sqrt{-\hat{g}}\Big[R(\hat{g},\hat{\Gamma})-\epsilon_{\text{Palatini}}\hat{g}^{\mu\nu}\hat{\nabla}_\mu\mathcal{I}^n_E\hat{\nabla}_\nu\mathcal{I}^n_E-\mathcal{I}^n_2\Big]+S_{\text{matter}}\Big[(\mathcal{I}^n_1)^{\frac{-2}{n-2}}\hat{g}_{\mu\nu},\chi\Big], \label{einstein}
	 \end{equation}
	 where $\mathcal{I}^n_1$ and $\mathcal{I}^n_2$ are functions of the invariant $\mathcal{I}^n_E$. 

Let us notice that if the invariant $\mathcal{I}^{n}_E$ vanishes, the scalar field has no dynamics, as the kinetic term is not present in the Lagrangian. In this case, the invariant $\mathcal{I}^{n}_2$ can be thought of as a function of the invariant $\mathcal{I}^{n}_1$ (the case in which $\mathcal{I}^{n}_E=0$ and $\mathcal{I}^{n}_2=0$ will not be considered, as such a theory is ill-posed). Regardless of which invariant will play the role of the scalar field, at the level of field equation the relation between the scalar field and the remaining fields will be purely algebraic, so that no additional physical degree of freedom will correspond to the extra scalar field included in the action. Since the transformation group acts always in a self-consistent way, this property must hold for all conformally related frames, for which $\mathcal{I}^{n}_E=0$. This is the case when $\epsilon_\text{Palatini}=0$ in the Einstein frame, thence all theories located on its orbit have no additional physical degree of freedom due to the presence of the scalar field. Moreover, at the level of the action functonal, a given theory may look as if it featured a dynamical scalar field (e.g. when $\mathcal{B}\neq0$, $\mathcal{C}_1\neq0$ and $\mathcal{C}_2\neq0$) but in fact it would be just an artifact of poorly chosen independent variables (metric and connection).  

As it can be seen, it is possible to find out a short cut  passage  from the complicated general action functional given by (\ref{action}) to a surprisingly simple and familiar form written above without using the group transformation rules. In the new frame, the scalar field is coupled only to matter part of the Lagrangian, which means that the Principle of Equivalence does not hold any more. The gravitational part is now free of terms $\mathcal{C}_1$ and $\mathcal{C}_2$, which were difficult to handle due to their coupling to the non-metricity tensors. Also, the kinetic coupling $\mathcal{B}$ is now equal to $\epsilon_{\text{Palatini}}$, leading to a further simplification of the field equations. 
	
	Variation with respect to all dynamical variables (assuming non-vanishing invariant $\mathcal{I}^n_E$) gives the following field equations:
		\begin{subequations}
			\begin{align}
			& \delta\hat{g}: \hat{G}_{\mu\nu}=\kappa^2\hat{T}_{\mu\nu}+\epsilon_{\text{Palatini}}\hat{\nabla}_\alpha\mathcal{I}^n_E\hat{\nabla}_\beta\mathcal{I}^n_E\Big(\delta^\alpha_\mu\delta^\beta_\nu-\frac{1}{2}\hat{g}^{\alpha\beta}\hat{g}_{\mu\nu}\Big)-\frac{1}{2}\hat{g}_{\mu\nu}\mathcal{I}^n_2, \\
			& \delta\hat{\Gamma}: \hat{\nabla}_\lambda\big(\sqrt{-\hat{g}}\:\hat{g}^{\mu\nu}\big)=0, \label{fe3}\\
			& \delta\mathcal{I}_3: 2\epsilon_\text{Palatini}\hat{\Box}\mathcal{I}^n_E-\frac{d\mathcal{I}^n_2}{d\mathcal{I}^n_E}=\kappa^2\frac{2-n}{2}\frac{1}{\mathcal{I}^n_1}\frac{d\mathcal{I}^n_1}{d\mathcal{I}^n_E}\hat{T}. 
			\end{align}
		\end{subequations}
If we consider the second equation, we immediately recognize the well-known relation between connection and metric tensor: if a connection is symmetric and the covariant derivative of the metric multiplied by its determinant vanishes, then the connection is necessarily Levi-Civita with respect to the metric. This shows an interesting result: after writing the action functional in terms of invariants, the initially independent invariant connection becomes Levi-Civita with respect to the invariant metric $\hat{g}_{\mu\nu}$. Consequently, the curvature scalar also depends on the metric. Apart from the presence of scalar field in the matter part of the action functional, this suggests that the Einstein  frame is supposedly the simplest.
		
			Alternatively, we can express the action functional in terms of the invariant metric $\tilde{g}_{\mu\nu}=e^{2\alpha(\Phi)}g_{\mu\nu}$, and the invariant linear connection $\hat{\Gamma}^{\alpha}_{\mu\nu}$. Also, the invariant $\mathcal{I}^n_1$ shall now play role of the scalar field. This will give us an action functional cast in a Jordan  frame:
			\begin{equation}
			S[\tilde{g}_{\mu\nu},\hat{\Gamma}^\alpha_{\mu\nu},\mathcal{I}^n_1]=\frac{1}{2\kappa^2}\int_{\Omega}d^nx\sqrt{-\tilde{g}}\Big[\mathcal{I}^n_1\hat{R}(\tilde{g},\hat{\Gamma})-\tilde{g}^{\mu\nu}\mathcal{I}^n_1\left(\frac{d\mathcal{I}^n_J}{d\mathcal{I}^n_1}\right)^2\hat{\nabla}_\mu\mathcal{I}^n_1\hat{\nabla}_\nu\mathcal{I}^n_1-\mathcal{I}^n_3\Big]+S_\text{matter}[\tilde{g}_{\mu\nu},\chi]. \label{jordan}
			\end{equation}

			For simplicity, we introduced another invariant, $\mathcal{I}^n_3$, defined in the following way: 
			$$\mathcal{I}^n_3=(\mathcal{I}_1^n)^{\frac{n}{n-2}}\mathcal{I}^n_2,$$
			denoting a modified potential.
			
			Let us now obtain equations of motion for the theory. Variation with respect to all three dynamical variables yields the following formulae:
			\begin{subequations}
				\begin{align}
				& \delta\tilde{g}: \hat{G}_{\mu\nu}(\tilde{g},\hat{\Gamma})=\frac{\kappa^2}{\mathcal{I}^n_1}\tilde{T}_{\mu\nu} +\Big(\frac{d\mathcal{I}^n_J}{d\mathcal{I}^n_1}\Big)^2\hat{\nabla}_\alpha\mathcal{I}^n_1\hat{\nabla}_\beta\mathcal{I}^n_1\big(\delta^\alpha_\mu\delta^\beta_\nu-\frac{1}{2}\tilde{g}_{\mu\nu}\tilde{g}^{\alpha\beta}\big)-\frac{1}{2}\tilde{g}_{\mu\nu}\frac{\mathcal{I}^n_3}{\mathcal{I}^n_1},\\
				& \delta\hat{\Gamma}: \hat{\nabla}_\alpha\big(\mathcal{I}^n_1\sqrt{-\tilde{g}}\tilde{g}^{\mu\nu}\big)=0, \\	\begin{split} &  \delta\mathcal{I}^n_1: \hat{R}(\tilde{g},\hat{\Gamma})-\tilde{g}^{\mu\nu}\hat{\nabla}_\mu\mathcal{I}^n_1\hat{\nabla}_\nu\mathcal{I}^n_1\Bigg[\Big(\frac{d\mathcal{I}^n_J}{d\mathcal{I}^n_1}\Big)^2+2\mathcal{I}^n_1\frac{d\mathcal{I}^n_J}{d\mathcal{I}^n_J}\frac{d^2\mathcal{I}^n_J}{d(\mathcal{I}^n_1)^2}\Bigg]-\frac{d\mathcal{I}^n_3}{d\mathcal{I}^n_1} \\
&\quad\quad\quad\quad+\frac{2}{\sqrt{-\tilde{g}}}\hat{\nabla}_\mu\Big(\sqrt{-\tilde{g}}\tilde{g}^{\mu\nu}\mathcal{I}^n_1\Big(\frac{d\mathcal{I}^n_J}{d\mathcal{I}^n_1}\Big)^2\hat{\nabla}_\nu\mathcal{I}^n_1\Big)=0.\end{split}
				\end{align}
			\end{subequations}
			Making use of the field equations, we can eliminate the independent invariant connection from (\ref{jordan}) and arrive at the action functional dependent on the metric and the scalar field only:
				\begin{equation}\label{invm4}
				\begin{split}
				S[\tilde{g}_{\mu\nu},\mathcal{I}^n_1]=&\frac{1}{2\kappa^2}\int_{\Omega}d^nx\sqrt{-\tilde{g}}\Bigg[\mathcal{I}^n_1\tilde{R}(\tilde{g})-\tilde{g}^{\mu\nu}\Bigg(\mathcal{I}^n_1\Big(\frac{d\mathcal{I}^n_J}{d\mathcal{I}^n_1}\Big)^2-\frac{n-1}{n-2}\frac{1}{\mathcal{I}^n_1}\Bigg)\hat{\nabla}_\mu\mathcal{I}^n_1\hat{\nabla}_\nu\mathcal{I}^n_1-\mathcal{I}^n_3\Bigg]\\&+S_\text{matter}[\tilde{g}_{\mu\nu},\chi]. 
				\end{split}
				\end{equation}
				
				For simplicity, let us introduce another invariant $\mathcal{I}^n_4$:
				$\mathcal{I}^n_4=\mathcal{I}^n_1\Big(\frac{d\mathcal{I}^n_J}{d\mathcal{I}^n_1}\Big)^2-\frac{n-1}{n-2}\frac{1}{\mathcal{I}^n_1}$. As it can be seen, if the invariant $\mathcal{I}^n_J$ is equal to zero, then $\mathcal{I}^n_4$ reduces to $-\frac{n-1}{n-2}\frac{1}{\mathcal{I}^n_1}$, so that the resultant theory in four dimensions is simply the standard Brans-Dicke theory with $\omega=-\frac{3}{2}$ and the modified self-interaction potential $\mathcal{I}^n_3$ added.

			\subsection{Scalar-tensor extension of  $F(\hat{R})$ gravity}
			By means of a simple transformation, it can be shown that $F(\hat{R})$ gravity is equivalent to special cases of  \cite{scap}, both in the metric and Palatini approach
\footnote{In this section $\hat R$ denotes, for short cut, Palatini-Ricci scalar, i.e. $\hat R= R(g,\Gamma)\equiv g^{\mu\nu}R_{\mu\nu}(\Gamma)$.}. This is achieved by a simple trick, as presented in the Appendix C. In fact, the metric $F(R)$ is equivalent to the Brans-Dicke (BD) theory with $\omega_{BD}=0$ (no kinetic term), while the Palatini $F(\hat{R})$ is equivalent to the Brans-Dicke theory with $\omega_{BD}=-\frac{n-1}{n-2}$ (with potential added to the Lagrangian in both cases and in $n$ dimensions). However, we may invert the problem and ask whether a given scalar-tensor gravity is equivalent to some $F(\hat R)$ theory (in mathematical, not physical sense). Answering this question might be much easier thanks to the introduction of invariant quantities, which are the same for different theories related to each other via conformal transformation. In order to find out whether two arbitrary theories can be linked by a transformation, we need to calculate the invariants and compare them. In this chapter, we will focus on $F(\hat{R})$ gravity and discuss conditions for equivalence with an S-T theory. 
			First, let us introduce the notion of Brans-Dicke theory in Palatini approach, which is a particular case of the Jordan frame (c.f. Definition V.2.)
			\begin{definition}
			\textbf{Brans-Dicke theory in Palatini approach} is given by the following action functional expressed in the Jordan frame:
			$$S[g_{\mu\nu},\Gamma^\alpha_{\mu\nu},\Psi]=\frac{1}{2\kappa^2}\int_{\Omega}d^nx \sqrt{-g}\Big(\Psi R(g,\Gamma)-\frac{\omega_\text{Palatini}}{\Psi}g^{\mu\nu}\nabla_\mu\Psi\nabla_\nu\Psi-\mathcal{U}(\Psi)\Big)+S_{\text{matter}}\left[g_{\mu\nu},\chi\right],$$
			
			with $\omega_\text{Palatini}=\text{const}$. 
			\end{definition}
	Brans-Dicke theory in the Palatini approach is not to be confused with the (original) BD theory in the metric approach, despite both of them having exactly the same functional form (see Appendix C). These theories are not physically equivalent, albeit one can show their mathematical equivalence. The proof goes as follows: using the fact that the BD theory in the Palatini approach is effectively metric, as it was proven in the previous section, one can express it the form analogous to (\ref{invm4}). Here, invariants $\mathcal{I}^n_1$ and $\mathcal{I}^n_2$ have exactly the same form, whereas the invariant $\mathcal{I}^n_J$ for a special choice of the function $\mathcal{B}$ is now:  \footnote{The sign $"-"$ corresponds to $\omega_\text{Palatini}<0$.}
	$$\mathcal{I}^n_J(\Psi)=\sqrt{\pm\: \omega_\text{Palatini}}\ln\left(\frac{\Psi}{\Psi_0}\right).$$
	Therefore, the (metric) action (\ref{invm4}) written for BD theory given initially in the Palatini approach, reads now as follows: 
	\begin{equation}
	S[g_{\mu\nu},\Psi]=\frac{1}{2\kappa^2}\int_{\Omega}d^nx \sqrt{-g}\Big(\Psi R(g)-\frac{\omega_\text{Palatini}-\frac{n-1}{n-2}}{\Psi}g^{\mu\nu}\nabla_\mu\Psi\nabla_\nu\Psi-\mathcal{U}(\Psi)\Big)+S_{\text{matter}}\left[g_{\mu\nu},\chi\right]. \label{e33}
	\end{equation}
	Let us observe that this action differs from (\ref{BDPn}), as the one written above is already evaluated on-shell, when the connection is Levi-Civita of the metric tensor. As it can be seen, when $\omega_\text{Palatini}=0$, the only difference is that the functions $\cc_1$ and $\cc_2$ do not vanish, so that they contribute to the field equation obtained from varying w.r.t. the metric and the independent connection. Therefore, the actions (\ref{e33}) and (\ref{BDPn}) are fully equivalent on-shell.
	
	The action written in the Einstein frame will have the following form (assuming $\omega_{\text{Palatini}}\neq 0$):
	
	\begin{equation}
	\begin{split}
	S[\bar{g}_{\mu\nu},\bar{\Psi}]=& \frac{1}{2\kappa^2}\int_{\Omega}d^nx \sqrt{-\bar{g}}\Big(R(\bar{g})\mp \bar{g}^{\mu\nu}\bar{\nabla}_\mu\bar{\Psi}\bar{\nabla}_\nu\bar{\Psi}-\bar{\mathcal{U}}(\bar{\Psi})\Big)\\
	& +S_{\text{matter}}\left[\exp\Big(-\frac{2}{n-2}\frac{\bar{\Psi}}{\sqrt{\pm \omega_{\text{Palatini}}}}\Big)\bar{g}_{\mu\nu},\chi\right].
	\end{split}
	\end{equation}	
	
	We may introduce the Brans-Dicke coefficient in the metric approach given in terms of
	\footnote{This result has been also found in \cite{Ulf,Ulf2}.}
	$$\omega_{BD}=\omega_\text{Palatini}-\frac{n-1}{n-2}.$$
Hence, the BD theory in the Palatini approach is equivalent to a BD in the metric formalism with the coefficient $\omega$ changed. 
Let us now ask a more general question: under what conditions is an arbitrary S-T theory equivalent to the BD theory by means of the transformation (\ref{e1})-(\ref{e3})? In order to resolve this issue, one needs to observe that for any theory to be equivalent to the BD, it must necessarily be expressible in the Jordan frame representation. In the transformed frame, one arrives at an action functional given by (\ref{jordan}). For this new action to describe a BD theory, it must possess the kinetic coupling of the form $\frac{\text{const}}{\bar{\Psi}}$, where $\bar{\Psi}$ is a function of the "old" scalar field $\phi$. Therefore, one might write the following equivalency condition:
\begin{equation}
\mathcal{I}^n_1(\phi)\left(\frac{d\mathcal{I}^n_J}{d\mathcal{I}^n_1}\right)^2=\pm\frac{\omega_\text{Palatini}}{\bar{\Psi}(\phi)}. \label{condition}
\end{equation}

From this point on, it will be very easy to give general conditions for mathematical equivalence between $F(\hat{R})$-Palatini  gravity and S-T theories. As it is shown, $F(\hat{R})$ gravity can be thought of as a (Palatini) Brans-Dicke theory with $\omega_\text{Palatini}= 0$ (or, equivalently, $\omega_{BD}=-\frac{n-1}{n-2}$, c.f. Appendix C). Therefore, in order to find out whether a given S-T theory in the Palatini approach arises from some $F(\hat{R})$ gravity, one needs to examine the condition (\ref{condition}) for $\omega_\text{Palatini}=0$. Such a condition is satisfied only when $\frac{d\mathcal{I}^n_J}{d\mathcal{I}^n_1}=0$, which means that (up to an additive constant) $\mathcal{I}^n_J=\mathcal{I}^n_E=0$. This reproduces the well-known result that there are only two physical degrees of freedom (graviton) in Palatini $F(\hat{R})$ theories of gravity \cite{olmo2009}.

When the equivalence is established, one may also wish to see what the exact form of the $F(\hat{R})$ function is. It is obvious that information about the $F(\hat{R})$ theory in the scalar-tensor representation is stored in the form of the potential defined as ${\cal U}(\Psi)=\Psi\:\Xi(\Psi)-F(\Xi(\Psi))$ (and $\hat{R}(\Psi)\equiv\Xi(\Psi)=\frac{d{\cal U}(\Psi)}{d\Psi}$) (see Appendix C). We find out that (assuming the coefficients defining the "old" frame - the one being subject to our inquiry - are $\{\mathcal{\bar{A}},\mathcal{\bar{B}},\mathcal{\bar{C}}_1,\mathcal{\bar{C}}_2,\mathcal{\bar{V}},\bar{\alpha}\}$, and the variables: $\{\bar{g},\bar{\Gamma},\bar{\Psi}\}$):
		 \begin{equation}
			 {\cal U}(\Psi)=\left(\mathcal{I}^n_1(\bar{\Psi}(\Psi))\right)^{\frac{n}{n-2}}\mathcal{I}^n_2(\bar{\Psi}(\Psi))\rightarrow \\\hat{R}(\Psi)\,,
			 \end{equation}
where
\begin{equation}		 
			\hat{R}(\Psi) =\frac{n}{n-2}\big(\mathcal{I}^n_1(\bar{\Psi}(\Psi))\big)^{\frac{2}{n-2}}\mathcal{I}^n_2(\bar{\Psi}(\Psi))+\left(\mathcal{I}^n_1(\bar{\Psi}(\Psi))\right)^{\frac{n}{n-2}}\frac{d}{d\Psi}\mathcal{I}^{n}_2(\bar{\Psi}(\Psi)).
			\end{equation}
			 The resulting equation is a non-linear differential equation of the first order, as $\Psi$ can be now identified with $\frac{dF}{d\hat{R}}$. Solving this equation will result in an exact form of the function $F(\hat{R})$.
			
\section{Conclusions}

			  In this paper, we have combined two frequently used ways of altering general relativity, Palatini variation and addition of a scalar field non-minimally coupled to the curvature, into a single theory of gravity. Our motivation for considering such coalescence of modifications of classical gravity was the lack of formalism of invariants defined for Palatini approach in S-T theories. Although the prevalent approach to the analysis of S-T theories is the metric one, the Palatini formalism has many interesting features to offer. 
			  
			  In the course of the paper, we placed special emphasis on the notion of conformal and almost-geodesic transformations, as it allows us to establish - under well-defined and strict conditions - mathematical equivalence between two different conformal frames. We did not aim to take a stand on the issue of which frame is the physical one; the main purpose of this paper was to obtain solution-equivalent classes of frames and introduce proper language enabling one to analyze the theory in a frame-independent manner. The first step to creating such language was to recognize that in case of the Palatini approach, one must transform the metric and the connection independently. Decoupling of metric from affine structure of spacetime influenced the action functional defined for a general S-T theory, devised to preserve its form under conformal change, enforcing us to add special terms linear in scalar field derivatives. These terms do not have any clear interpretation yet. 
			  
			  We singled out two frames most commonly used in the literature - Jordan and Einstein. Quantities behaving as invariants on the orbits of the two frames were also introduced and the role they play when comparing equivalent theories was discussed. In general, the theory possesses three degrees of freedom: one introduced by the scalar field, and the remaining two being a property of the metric. However, the independent scalar field turns out to be an auxiliary field in case the invariant $\mathcal{I}^n_E$ vanishes; then, the theory has only two degrees of freedom.
			  
			  It was discovered that there exists a subclass of conformal frames with $\mathcal{C}_1=\mathcal{C}_2=0$ fully analogous to the metric frames. In such frames, the (initially independent) connection is always Levi-Civita with respect to a metric $\bar{g}$ conformally related to the initial metric $g$. This class is invariant under the action of the subgroup $\gamma_2=\gamma_3=0$. 
			  
			  If a given theory has the same $\{\mathcal{A},\mathcal{B},\mathcal{V},\alpha\}$ functions both in the metric and Palatini approach, the latter one can be brought to the metric form using the property discussed above. The only difference between such two theories will be the exact form of the kinetic coupling $\mathcal{B}$; in the metric formalism resulting from a prior Palatini frame, the coupling will take on the form $\mathcal{B}-\frac{n-1}{n-2}\frac{1}{\Phi}$. This fact allowed us to establish a correspondence between the Brans-Dicke theories in the metric and Palatini formalism. 
			  
			  It was also shown that for an arbitrary S-T theory in the Palatini approach there always exists a unique transformation defined for the connection such that it renders the theory effectively metric. This useful property allows us to analyze a specific theory within the metric formalism. 
			  
			  Finally, $F(\hat{R})$ theories were analyzed using the language of invariants. We made use of the well-established equivalence of these theories to S-T gravity - to the Brans-Dicke theory, to be precise. Invariants made it possible for us to address an issue of the relation between S-T and $F(\hat{R})$, namely, we identified cases in which those two theories could be related by the transformation (\ref{e1})-(\ref{e3}), meaning that they are mathematically equivalent. It was discovered that the coefficients $\{\mathcal{A},\mathcal{B},\mathcal{C}_1,\mathcal{C}_2,\mathcal{V},\alpha\}$, which characterize a specific S-T theory, must fulfil certain relations (given by (\ref{condition})) in order for the theory to be equivalent to $F(\hat{R})$ gravity in the Palatini approach. Furthermore, because the metric and the Palatini formalisms always give two non-equivalent theories, if a given scalar-tensor theory results from some $F(R)$ theory, it cannot simultaneously be derived from both the metric and the Palatini $F(R)$. 

The main aim of this paper was to introduce a new class of scalar-tensor theories of gravity and analyze some of its mathematical properties. Due to its introductory nature, it focuses on the formal aspects of the theory, with a special emphasis put on self-consistency conditions, and lacks direct physical applications.  Also, due to adopting the Palatini approach and adding more degrees of freedom into the theory, it will be straightforward to include torsion and/or disformal transformations in order to investigate theirs impact on self-consistency of the theory. Analysis of real-world phenomena will be carried out in the forthcoming papers. In order to find out whether the predictions of the theory are in agreement with experiment, we plan on computing the post-Newtonian parameters in the first place.  Furthermore, topics to be covered in the future works will include cosmological applications (cf. \cite{Bor:2016,ssb}), F(R) theories with non-minimal curvature coupling (see e.g. \cite{Allem:2005,Bor:2011}), the appearance of ghosts and tachions.

\subsection*{Acknowledgments}
We are grateful to Ulf Lindstr\"{o}m for helpful comments concerning his earlier papers on the subject.
This research was supported by Polish National Science Center (NCN), project
UMO-2017/27/B/ST2/01902.


	\begin{appendices}
\renewcommand{\theequation}{A.\arabic{equation}}
\setcounter{equation}{0}
 
		\section{Metric scalar-tensor gravity}
	For the sake of completeness we recall the formalism introduced in  \cite{flan,jarv}, slightly generalized to arbitrary dimension $n>2$ \cite{Karam}.  The action functional is:
			\begin{equation}
			\begin{split}
			S[g_{\mu\nu},\Phi]&=\frac{1}{2\kappa^2}\int_{\Omega}d^nx \sqrt{-g} \Big(\mathcal{A}(\Phi) R(g)-\mathcal{B}(\Phi) g^{\mu\nu} \nabla_\mu \Phi \nabla_\nu \Phi -\mathcal{V}(\Phi)\Big) \\
			& + S_\text{matter}\Big[e^{2\alpha(\Phi)}g_{\mu\nu},\chi\Big].
			\end{split}
			\end{equation}
			Varying the action functional with respect to the metric tensor yields:
			
			\begin{equation}
			\begin{split}
			&\mathcal{A}(\Phi)G_{\mu\nu}+\Big(\frac{1}{2}\mathcal{B}+\mathcal{A}''\Big)g_{\mu\nu}g^{\alpha\beta}\nabla_\alpha\Phi\nabla_\beta\Phi-\big(\mathcal{B}+\mathcal{A}''\big)\nabla_\mu\Phi\nabla_\nu\Phi+\mathcal{A}'\big(g_{\mu\nu}\Box-\nabla_\mu\nabla_\nu\big)\Phi - \\
			&+\frac{1}{2}g_{\mu\nu}\mathcal{V}-\kappa^2T_{\mu\nu}=0,
			\end{split} \label{eom}
			\end{equation}
			
			\noindent with the standard definition of the energy-momentum tensor, $T_{\mu\nu}=\frac{2}{\sqrt{-g}}\frac{\partial (\sqrt{-g}\mathcal{L}_m)}{\partial g^{\mu\nu}}$, $\mathcal{L}_m$ being Lagrangian for matter. Variation with respect to the scalar field gives:
			
			\begin{equation}
			R\mathcal{A}'+\mathcal{B}'g^{\mu\nu}\nabla_\mu\Phi\nabla_\nu\Phi+2\mathcal{B}\:\Box\Phi-\mathcal{V}'+2\kappa^2\alpha'T=0.
			\end{equation}
			
			The scalar field is sourced by the trace of energy-momentum tensor. The continuity equation takes the following form:
			
			\begin{equation}
			\nabla^\nu T_{\mu\nu}=\frac{d\alpha(\Phi)}{d\Phi}T\:\nabla_\mu \Phi.
			\end{equation}
			
			Two of the four arbitrary functions can be fixed by means of a conformal change accompanied by a redefinition of the scalar field:
			\footnote{This implies that the Levi-Civita connection undergoes the Weyl transformation $\bar\Gamma^\alpha_{\mu\nu}=\Gamma^\alpha_{\mu\nu}+2 \delta^\alpha_{(\mu}\partial_{\nu)}\gamma_2(\Phi)-g_{\mu\nu} g^{\alpha\beta}\partial_\beta\gamma_2(\Phi) $. }
			\begin{subequations} 
				\begin{align}
			&\bar g_{\mu\nu}=e^{2\gamma(\Phi)}g_{\mu\nu}, \label{ct1}\\
			&\bar \Phi=f(\Phi).  \label{ct2}
			\end{align}
			\end{subequations}
			It is generally assumed that the first and second derivatives of $\bar{\gamma}$ exist. Moreover, the Jacobian of the transformation is allowed to be singular at some isolated point \cite{jarv}. 
			
			If we plug the redefined scalar field and metric tensor back in the action functional, make use of the transformation relations and neglect boundary terms arising while integrating by parts, we end up with the action written in a different conformal frame, with the barred dynamical variables. In order for the Lagrangian to retain its form, the coefficients must transform in the following way (for the notational convention see next Section):
			\begin{subequations} \label{eqns7}
				\begin{align}
				\bar{\mathcal{A}}(\bar{\Phi})&=e^{(n-2)\check{\gamma}(\bar{\Phi})}\mathcal{A}(\check{f}(\bar{\Phi})), \\
				\bar{\mathcal{B}}(\bar{\Phi})&=e^{(n-2)\check{\gamma}(\bar{\Phi})}\Bigg(\Big(\frac{d\Phi}{d\bar{\Phi}}\Big)^2\mathcal{B}(\check{f}(\bar{\Phi}))-(n-1)(n-2)\Big(\frac{d\check{\gamma}}{d\bar{\Phi}}\Big)^2\mathcal{A}(\check{f}(\bar{\Phi}))-2(n-1)\frac{d\check{\gamma}}{d\bar{\Phi}}\frac{d\mathcal{A}}{d\Phi}\frac{d\Phi}{d\bar{\Phi}}\Bigg),\\
				\bar{\mathcal{V}}(\bar{\Phi})&=e^{n\check{\gamma}(\bar{\Phi})}\mathcal{V}(\check{f}(\bar{\Phi})),\\
				\bar{\alpha}(\bar{\Phi})&=\alpha(\check{f}(\bar{\Phi}))+\check{\gamma}(\bar{\Phi}).
				\end{align}
			\end{subequations}
			The transformation relations suggest that the conditions imposed on $\mathcal{A}$ and $\mathcal{V}$ are satisfied in any conformal frame. In particular, if the potential vanishes in one conformal frame, then it is equal to zero in all related conformal frames. Let us also make a comment regarding the nomenclature: choosing the functions defining the conformal transformation will be called "fixing the frame", while setting the remaining two coefficients will be equivalent to choosing a particular theory. 
			
			It is possible to define the following invariants:
			\begin{enumerate}
				\item $\mathcal{I}_1(\Phi)=\frac{\mathcal{A}(\Phi)}{e^{(n-2)\alpha(\Phi)}},$ 
				
				\item $\mathcal{I}_2(\Phi)=\frac{\mathcal{V}(\Phi)}{(\mathcal{A}(\Phi))^{\frac{n}{n-2}}},$
				
				\item $\frac{d\mathcal{I}_3(\Phi)}{d\Phi}= \sqrt{\pm\frac{(n-2)\mathcal{A}(\Phi)\mathcal{B}(\Phi)+(n-1)(\mathcal{A}'(\Phi))^2}{n\mathcal{A}^2(\Phi)}}$. 
			\end{enumerate}
			
			Alongside the invariants defined above, we may introduce invariant metrics, remaining unchanged under a conformal transformation:
		\begin{subequations} 
				\begin{align}
			\hat{g}_{\mu\nu}&:=(\mathcal{A}(\Phi))^{\frac{2}{n-2}} g_{\mu\nu},\\
			\tilde{g}_{\mu\nu}&:=e^{2\alpha(\Phi)}g_{\mu\nu}
			\end{align}
			\end{subequations} 
(invariance of this metric follows from transformation properties of $\mathcal{A}$, $e^{2\alpha(\Phi)}$ and the metric tensor $g_{\mu\nu}$). Invariance of the metric tensor simply means that if observers of different conformal frames being related to each other by means of (\ref{ct1}) and (\ref{ct2}) agree on using one of the above metrics, then the distances measured by them will be the same. 
\renewcommand{\theequation}{B.\arabic{equation}}
\setcounter{equation}{0}

\section{Transformation groups and their consistent actions}	
Consider diffeomorphism group of real line $\mathtt{Diff}(\mathbb{R})$ 
\footnote{Since $f^\prime\neq 0$ one can also consider a subgroup  $f^\prime>0$.} with multiplcation given by the composition law. It can be extended (as a semi-direct product) by an arbitrary number of functions $\gamma_i\in C^1(\mathbb{R})$ acting as generalized translations. The resulting group  with the multiplication law 
\begin{equation}\label{b1}
(\bar f, \bar\gamma_1,\dots, \bar\gamma_r)\circ(f, \gamma_1,\dots, \gamma_r)=(\bar f\circ f, \bar\gamma_1+\gamma_1\circ \bar f^{-1},\dots, \bar\gamma_r+\gamma_r\circ \bar f^{-1})\equiv
(\bar{\bar{f}}, \bar{\bar\gamma}_1,\dots, \bar{\bar\gamma}_r)
\end{equation}
is denoted as $\mathtt{Diff}^{(r)}(\mathbb{R})$. The inverse element has the form
\begin{equation}\label{b2}
(f, \gamma_1,\dots, \gamma_r)^{-1}=(f^{-1}, -\gamma_1\circ f,\dots, -\gamma_r\circ f)\equiv
(\check f, \check\gamma_1,\dots, \check\gamma_r).
\end{equation}
Such group  admits  several subgroups, e.g. $\mathtt{Diff}^{(r)}(\mathbb{R})\subset \mathtt{Diff}^{(s)}(\mathbb{R})$ for
$r<s$ or by imposing some linear relations between the generators $\gamma_i$, e.g. $\gamma_1=-\gamma_2$.

Here we are interested in  $\mathtt{Diff}^{(r)}(\mathbb{R})$-spaces representing some differential-geometric structures on a manifold. In the case of Riemannian metric and a scalar field $(g_{\mu\nu}, \Phi)$
this action of  $\mathtt{Diff}^{(1)}(\mathbb{R})$ has the form (c.f. (\ref{ct1})-(\ref{ct2}))
$$(f,\gamma^{})\rhd (g_{\mu\nu}, \Phi)= (\exp{(2\gamma(\Phi))} g_{\mu\nu}, f\circ\Phi)\equiv (\bar g_{\mu\nu}, \bar\Phi).$$
One can notice that  $\gamma = \textit{const}$ acts trivially by rescaling the metric by a numerical constant. This action obeys consistency condition: the result of consecutive actions
\begin{equation}\label{b3}
(\bar f, \bar\gamma)\rhd[(f,\gamma^{})\rhd (g_{\mu\nu}, \Phi)]\equiv (\bar f, \bar\gamma)\rhd
 (\bar{ g}_{\mu\nu}, \bar{\Phi})\end{equation} 
 must be the same as an action by their composition
\begin{equation}\label{b4}
[(\bar f, \bar\gamma)\circ(f,\gamma^{})]\rhd (g_{\mu\nu}, \Phi)]
 \equiv (\bar{\bar f}, \bar{\bar\gamma})\rhd ( g_{\mu\nu}, \Phi) 
 \equiv   (\bar{\bar{ g}}_{\mu\nu}, \bar{\bar{\Phi}})\,. \end{equation}
Similarly, the group  $\mathtt{Diff}^{(3)}(\mathbb{R})$ acts, in the consistent way by (\ref{e1})-(\ref{e3}), onto the collection of dynamical variables  $(g, \Gamma, \Phi)$ of the S-T Palatini theory, which represent independent variables. The kernel of this action  consists of constant functions ($\gamma_i=\text{const}_i$). In particular, one can reduce this group to a subgroup 
isomorphic to $\mathtt{Diff}^{(2)}(\mathbb{R})$
containing, e.g.  projective or Weyl transformation of the connection, i.e. $\gamma_3=0$, resp. $\gamma_2=\gamma_3$.
Strict Weyl transformations can be defined by the condition $\gamma_1=\gamma_2=\gamma_3$. The subgroup of Weyl transformations is isomorphic to $\mathtt{Diff}^{(1)}(\mathbb{R})$. In this sense  the action  (\ref{e1}-\ref{e3}) of $\mathtt{Diff}^{(3)}(\mathbb{R})$ generalizes (\ref{ct1})-(\ref{ct2}) of $\mathtt{Diff}^{(1)}(\mathbb{R})$.
 
One can directly  check that the action (\ref{ct1})-(\ref{ct2}) induces a consistent action on the space of metric frames represented by the  the collection of functions of one-real variable 		$\{\ca,\cb, \cv, \alpha\}$ composed with the scalar field $\Phi$ (see eqs. \ref{eqns7}).
Similarly, the induced action  (\ref{t1})-(\ref{t6}) of $\mathtt{Diff}^{(3)}(\mathbb{R})$ on the collection of functions representing Palatini frames (dependent variable) $\{\ca, \cb, \cc_1, \cc_2, \cv, \alpha\}$ is also consistent, which can be demonstrated directly by composing two subsequent generalized conformal transformations.  \footnote{More general action with the gradient field $\partial_\alpha\Phi$ replaced by an arbitrary one form will be considered elsewhere.}

\renewcommand{\theequation}{C.\arabic{equation}}
\setcounter{equation}{0} 

 \section{From $F(R)$ to scalar-tensor gravity}
		
In this subsection we review  
the traditional approach to both metric as well as Palatini  $F(R)$-gravity.  As it is well-known, in both cases,  $F(R)$- gravity is dynamically equivalent to so-called Brans-Dicke (BD) theories. Original  BD is a metric S-T theory determined by the gravitational action:
\begin{equation}\label{actionBD}
 S_{BD}(g_{\mu\nu}) =\frac{1}{2\kappa^2}\int_\Omega\mathrm{d}^nx\sqrt{-g}\left(\Phi R -\frac{\omega_{BD}}{\Phi}\partial_\mu\Phi\partial^\mu\Phi- U(\Phi) \right),
\end{equation}		
where BD parameter  $\omega_{BD}\in \mathbb{R}$ and $U(\Phi)$ denotes self-interaction potential.		
As we have already pointed out,  mathematically equivalent theories are not physically equivalent. 
		
Consider the action of minimally coupled $F(R)$-gravity
	\begin{equation}\label{Paction}
 S_F(g_{\mu\nu}, .)=\frac{1}{2\kappa^2}\int_\Omega\mathrm{d}^nx\sqrt{-g}F( R)+ S_\text{matter}(g_{\mu\nu},\chi),
\end{equation}
where $F(R)$ is a function either a  Ricci or a Palatini scalar. The matter part of the action $S_\text{matter}$ is assumed metric-dependent (independent of the connection).	
 In both cases the action (\ref{Paction}) is dynamically equivalent to the constraint system with linear gravitational Lagrangian \footnote{One should stress that Palatini $F(R)$-gravity is not dynamically equivalent to  metric one with the same function $F(R)$.}
\begin{equation}\label{action1}
 S(g_{\mu\nu}, . , \Xi)=\frac{1}{2\kappa^2}\int_\Omega\mathrm{d}^nx\sqrt{-g}\left(F^\prime(\Xi)( R-\Xi) + F(\Xi) \right) + S_\text{matter}(g_{\mu\nu},\chi).
\end{equation}
Introducing further a scalar field $\Phi=F'(\Xi)$ and taking into account the constraint equation $\Xi=R$, one arrives to the dynamically equivalent S-T action  with non-dynamical scalar field
\begin{equation}\label{actionP}
 S(g_{\mu\nu}, . ,\Phi)=\frac{1}{2\kappa^2}\int_\Omega\mathrm{d}^nx\sqrt{-g}\left(\Phi R - U_F(\Phi) \right)+ S_\text{matter}(g_{\mu\nu},\chi) 
\end{equation}
either in metric or Palatini case.  The self-interaction  potential  $U_F(\Phi)$   is induced from the function $F(R)$  by the following formula 
\begin{equation}\label{PotentialP}
 U_F(\Phi)\equiv \Xi(\Phi)\Phi-F(\Xi(\Phi))\,,
\end{equation}
where $\Phi = \frac{d F(\Xi)}{d\Xi}$ and $ R\equiv \Xi = \frac{d U_F(\Phi)}{d\Phi}$ \footnote{One can observe that the trivial, i.e. constant,  potential $U(\Phi)$ corresponds to the linear Lagrangian $F( R)=R-2\Lambda$. More generally, for a given $F$ the potential $U_F$ is a (singular) solution of the Clairaut's differential equation: $U_F(\Phi)=\Phi \frac{d U_F}{d\Phi} -F(\frac{d U_F}{d\Phi})$.}. Thus, in the metric case, the action (\ref{actionP}) represents Brans-Dicke theory with the Brans-Dicke scalar $\omega_{BD}=0$ minimally coupled to the matter field. 

Palatini variation of this action provides
\footnote{It also corresponds to the Palatini Brans-Dicke theory,  in a sense of Definition V.3, with $\omega_\text{Palatini}=0$.}
\begin{subequations}	
	\begin{align}
	\label{EOM_P}
	\Phi\left( R_{(\mu\nu)}(\Gamma) - \frac{1}{2} g_{\mu\nu} g^{\alpha\beta} R_{\alpha\beta}(\Gamma)\right) &  +{1\over 2} g_{\mu\nu} U_F(\Phi) - \kappa^2 T_{\mu\nu} = 0\,,\\
	\label{EOM_connectP}
	& \nabla^\Gamma_\lambda(\sqrt{-g}\Phi g^{\mu\nu})=0\,,\\
	\label{EOM_scalar_field_P}
	  g^{\alpha\beta} &R_{\alpha\beta}(\Gamma)  -  U_F^\prime(\Phi) =0\,.
	\end{align}
\end{subequations}
The last equation due to the constraint $g^{\alpha\beta} R_{\alpha\beta}(\Gamma)=\Xi= U_F^\prime(\Phi)$ is automatically satisfied. The middle equation (\ref{EOM_connectP}) implies that the connection $\Gamma$ is a metric connection for the new metric $\bar g_{\mu\nu}=\Phi^{\frac{2}{n-2}} g_{\mu\nu}$.
 
Now, we can switch  from the original connection $\Gamma^\lambda_{\mu\nu}$ to Levi-Civita connection of the original metric $g_{\mu\nu}$ by performing Weyl transformation of the connection  (without  changing the metric), i.e. with the parameters $\gamma_1=0, \gamma_2=\gamma_3=-\frac{\ln \Phi}{n-2}$. As a result one gets the minimally coupled metric theory with the following action: 
\begin{equation}\label{BDPn}
\begin{split}
&S_{BD}(g_{\mu\nu})=\frac{1}{2\kappa^2}\int_\Omega\mathrm{d}^nx\sqrt{-g}\left(\Phi R +\frac{n-1}{(n-2)\Phi}\partial_\mu\Phi\partial^\mu\Phi+A^\mu_1\partial_\mu\Phi + A^\mu_2\partial_\mu\Phi - U_F(\Phi) \right)\\
&+S_\text{matter}(g_{\mu\nu},\chi).
\end{split}
\end{equation}
In this case, a kinematical  part of the scalar field does not vanish from the Lagrangian (\ref{actionP}). This action is clearly not represented in the Jordan frame, as the coefficients $\cc_1=\cc_2$ do not vanish, but are equal to $-1$ instead. However, this theory turns out to be metric on-shell, i.e. the connection solving EOM is Levi-Civita w.r.t. the initial metric tensor, even though the action contains the terms which have not been taken into account so far. Also, despite the presence of kinetic term for the scalar field, it is not dynamical, as the invariant $\mathcal{I}^n_J$ vanishes.

In order to obtain the so-called Einstein frame it is enough now to choose $\gamma=\gamma_1=\frac{\ln\Phi}{n-2}$ and to apply it to the action (\ref{actionP}).
In the metric case we obtain  non-minimally-coupled theory with the action
\begin{equation}\label{E-m}
 \tilde S_{}(g_{\mu\nu})=\frac{1}{2\kappa^2}\int_\Omega\mathrm{d}^nx\sqrt{-g}\left(R -\frac{n-1}{(n-2)\Phi^2}\partial_\mu\Phi\partial^\mu\Phi- \bar U_F(\Phi) \right) +S_\text{matter}(\Phi^{-\frac{2}{n-2}}g_{\mu\nu},\chi),
\end{equation}
where the potential $U_F$ is now replaced by $\bar U_F:= \frac{U_F}{\Phi^{\frac{n}{n-2}}}$.
Performing field re-definition by introducing new scalar field $\bar\Phi=\sqrt{\frac{n-1}{n-2}}\ln\Phi$ one can arrive at the action with the parameter $\bar\cb=1$:
\begin{equation}\label{E-m2}
 S_{E}(g_{\mu\nu})=\frac{1}{2\kappa^2}\int_\Omega\mathrm{d}^nx\sqrt{-g}\left(R -\partial_\mu\bar\Phi\partial^\mu\bar\Phi- \bar U_F(e^{\sqrt{\frac{n-2}{n-1}}\bar\Phi}) \right) +S_\text{matter}(e^{-\sqrt{\frac{4}{(n-1)(n-2)}}\bar\Phi}g_{\mu\nu},\chi).
\end{equation}

Palatini case leads to non-minimally coupled metric theory without kinetic term for the scalar field
\begin{equation}\label{E-P}
 S_{EP}(g_{\mu\nu})=\frac{1}{2\kappa^2}\int_\Omega\mathrm{d}^nx\sqrt{-g}\left(R - \bar U_F(\Phi) \right)+ S_\text{matter}(\Phi^{-\frac{2}{n-2}}g_{\mu\nu},\chi)\,,
\end{equation}
which agrees with the Einstein frame Definition V.1.

We see that in both cases the matter part bears the same non-minimal coupling between the metric and the matter, and that the potential $U_F$ is modified in the same way.

Remark: Assuming non-minimal coupling in $F(R)$ theory (as e.g. in \cite{Allem:2005})  one would be able  to reach minimal coupling in the Einstein frame.

\renewcommand{\theequation}{D.\arabic{equation}}
\setcounter{equation}{0}
 
		\section{Almost-geodesic mappings}
		The content of this Appendix was written based on \cite{ber}, \cite{bere}, \cite{diff}. In order to introduce the notion of an almost geodesic mapping, one must define the following concept:
		
		\begin{definition}
			A curve $\gamma$ in a space endowed with an affine connection $A_n$ is called \textit{almost geodesic} if there exists a two-dimensional parallel distribution along $\gamma$, to which the tangent vector of this curve belongs at every point
		\end{definition}
		
		An almost geodesic mapping is defined as follows:
		
		\begin{definition}
			A diffeomorphism $f:\:A_n\rightarrow\bar{A}_n$ is called \textit{an almost geodesic mapping} if every geodesic curve of $A_n$ is transformed by $f$ into an almost geodesic curve of $\bar{A}_n$.
		\end{definition}

		In order for $f$ to be almost geodesic, the condition given below must be satisfied:
		
		\begin{theorem}
			A mapping $f: \:A_n\rightarrow\bar{A}_n$ is almost geodesic iff in a common coordinate system $\{x^\alpha\}_{\alpha=1}^n$, the connection deformation tensor $P^\alpha_{\mu\nu}:=\bar{\Gamma}^{\alpha}_{\mu\nu}-\Gamma^\alpha_{\mu\nu}$ satisfies the relation:
			\begin{equation}
			A^\alpha_{\mu\nu\beta}\lambda^\mu\lambda^\nu\lambda^\beta=a(x,\lambda)P^\alpha_{\mu\nu}\lambda^\mu\lambda^\nu+b(x,\lambda)\lambda^\alpha,
			\end{equation}
			where $A^\alpha_{\mu\nu\beta}=\nabla^\Gamma_\beta P^\alpha_{\mu\nu}+P^\sigma_{\mu\nu}P^\alpha_{\sigma\beta}$, $\Gamma^\alpha_{\mu\nu}$ is an affine connection on $A_n$ (and, analogously, $\bar{\Gamma}^\alpha_{\mu\nu}$ is a connection on $\bar{A}_n$), $\lambda^\alpha$ is any vector, $a$ and $b$ are some functions of $x^\alpha$ and $\lambda^\alpha$. The covariant derivative $\nabla^\Gamma$ is defined with respect to the connection $\Gamma^\alpha_{\mu\nu}$.
		\end{theorem}
		
		There are three types of almost geodesic mappings, as distinguished by N. S. Sinyukov \cite{sin}, \cite{siny}:
		
		\begin{enumerate}
			\item \textbf{type $\pi_1$:} \begin{equation}
			\nabla^\Gamma_{(\beta}P^\alpha_{\mu\nu)}+P^\sigma_{(\mu\nu}P^\alpha_{\beta)\sigma}=\delta^\alpha_{(\mu}a_{\nu\beta)}+b_{(\mu}P^\alpha_{\beta\nu)},
			\end{equation}
			where $a_{\mu\nu}$ and $b_\mu$ are tensors;
			\item \textbf{type $\pi_2$: } \begin{subequations}
				\begin{align}
				& P^\alpha_{\mu\nu}=\delta^\alpha_{(\mu}\psi_{\nu)}+F^\alpha_{(\mu}\phi_{\nu)}, \\
				& \nabla^\Gamma_{(\mu}F^\alpha_{\nu)}+F^\alpha_\sigma F^\sigma_{(\mu}\phi_{\nu)}=\delta^\alpha_{(\mu}\omega_{\nu)}+F^\alpha_{(\mu}\sigma_{\nu)},
				\end{align}
			\end{subequations}
			where $F^\alpha_\mu$ is a tensor of type $(1,1)$ and $\psi_\mu, \phi_\mu, \omega_\mu, \sigma_\mu$ are covectors;
			\item \textbf{type $\pi_3$: } \begin{subequations}
				\begin{align}
				& P^\alpha_{\mu\nu}=\delta^\alpha_{(\mu}\psi_{\nu)}+\phi^\alpha\omega_{\mu\nu}, \\
				& \nabla^\Gamma_\mu\phi^\alpha=\rho \delta^\alpha_\mu+\phi^\alpha a_\mu,
				\end{align}
			\end{subequations}
			where $\alpha_\mu, a_\mu$ are covectors, $\phi^\alpha$ is a vector, $\omega_{\mu\nu}$ is a symmetric tensor and $\rho$ is a function.
		\end{enumerate}
		
	\end{appendices}

	\end{document}